\documentclass[10pt,journal,compsoc]{IEEEtran}
\pdfoutput=1

\usepackage{booktabs}
\usepackage{balance}
\usepackage[mathscr]{euscript}
\usepackage{verbatim}
\usepackage{amssymb, amsmath, color, graphicx, amsbsy, bm}

\usepackage{epsfig}
\usepackage{amsthm}

\usepackage{fixltx2e}
\usepackage{xcolor}

\allowdisplaybreaks[1]

\usepackage{bbm}
\usepackage{amssymb}
\usepackage{algorithmic}       
\usepackage{amsmath}
\usepackage{graphicx}
\usepackage{booktabs}
\usepackage[flushleft]{threeparttable}
\usepackage{multirow}
\usepackage{amsthm}
\usepackage{graphicx,booktabs,multirow}
\usepackage{stmaryrd}
\usepackage{multirow,url}
\usepackage{algorithm}
\usepackage{cite}
\usepackage{color}

\usepackage{xpatch}
\usepackage{subfloat}
\usepackage{dsfont}

\usepackage{subfigure}
\usepackage{multirow}
\usepackage{makecell}
\usepackage{float}

\def\T{\mathcal{T}}
\def\K{\mathcal{K}} 

\begin{document}
	
	\title{Evolutionary Multi-Objective Reinforcement Learning Based Trajectory Control and Task Offloading in UAV-Assisted Mobile Edge Computing}
	
	\author{Fuhong Song, Huanlai Xing, \textit{Member}, IEEE, Xinhan Wang, Shouxi Luo, \textit{Member}, IEEE, Penglin Dai, \textit{Member}, IEEE, Zhiwen Xiao, Bowen Zhao
			
		\IEEEcompsocitemizethanks{\IEEEcompsocthanksitem F. Song, H. Xing, X. Wang, S. Luo, P. Dai, B. Zhao are with the School of Computing and Artificial Intelligence, Southwest Jiaotong University, Chengdu 610031, China (E-mail: fhs@my.swjtu.edu.cn, hxx@home.swjtu.edu.cn, xhwang@my.swjtu.edu.cn,  sxluo@swjtu.edu.cn, penglindai@swjtu.edu.cn, cn16bz@icloud.com).
		
	\IEEEcompsocthanksitem Z. Xiao is with Southwest Jiaotong University, Chengdu 611756, China, and Chengdu University of Information Technology, Chengdu 610103, China  (Email: xiao1994zw@163.com).

}}
	
\IEEEtitleabstractindextext{
\begin{abstract}
This paper studies the trajectory control and task offloading (TCTO) problem in an unmanned aerial vehicle (UAV)-assisted mobile edge computing system, where a UAV flies along a planned trajectory to collect computation tasks from smart devices (SDs). We consider a scenario that SDs are not directly connected by the base station (BS) and the UAV has two roles to play: MEC server or wireless relay. The UAV makes task offloading decisions online, in which the collected tasks can be executed locally on the UAV or offloaded to the BS for remote processing. The TCTO problem involves multi-objective optimization as its objectives are to minimize the task delay and the UAV's energy consumption, and maximize the number of tasks collected by the UAV, simultaneously. This problem is challenging because the three objectives conflict with each other. The existing reinforcement learning (RL) algorithms, either single-objective RLs or single-policy multi-objective RLs, cannot well address the problem since they cannot output multiple policies for various preferences (i.e. weights) across objectives in a single run. This paper adapts the evolutionary multi-objective RL (EMORL), a multi-policy multi-objective RL, to the TCTO problem. This algorithm can output multiple optimal policies in just one run, each optimizing a certain preference. The simulation results demonstrate that the proposed algorithm can obtain more excellent nondominated policies by striking a balance between the three objectives regarding policy quality, compared with two evolutionary and two multi-policy RL algorithms.
\end{abstract}

}
	
\maketitle
	
\IEEEdisplaynontitleabstractindextext
	
\IEEEpeerreviewmaketitle
	

\section{Introduction}
With the rapid development of Internet-of-Things (IoT) technology, smart devices (SDs) play an essential role in various applications, such as object detectors for autonomous control, high definition cameras for intelligent grazing, and meteorological sensors for environmental monitoring \cite{wang2020optimal}. SDs can be deployed to monitor and collect data from areas of interest, thus providing new opportunities for emerging intelligent applications, e.g., industrial automation and smart city. These applications are usually computing-intensive, which results in dramatically increased demand for computing resources, posing a great challenge to SDs due to their limited computing resources and battery capacity \cite{song2022offloading}. 

The contradiction between computing-intensive applications and resource-constrained SDs creates a bottleneck when achieving satisfactory quality of experience (QoE) for end users. Fortunately, mobile edge computing (MEC) brings abundant computing resources to the edge of networks close to SDs \cite{mach2017mobile}. Under this paradigm, SDs can offload computing-intensive applications to nearby terrestrial base stations (BSs), which reduces the processing delay of applications and saves the energy consumption of SDs. Migrating these applications to BSs for execution are also referred to as computation offloading. Although the traditional BS-based MEC promotes computing-intensive applications in many fields, including computation and communication, MEC with BSs only may not always results in satisfactory computation offloading performance \cite{zhou2020deep}. A terrestrial BS has a fixed wireless communication coverage while users can be anywhere. It is not possible for a BS to connect to a user out of its coverage. Especially some BSs may be damaged by natural disasters or military attacks, causing computing resource scarcity and offloading performance degradation \cite{zhang2020energy}. How to provide users with on-demand computing services is one of the main challenges BS-based MEC networks face. Thanks to its high mobility and excellent maneuverability, unmanned aerial vehicle (UAV) has been applied to terrestrial networks for communication coverage extension and deployment efficiency improvement \cite{ning2021dynamic,liu2019uav}. Generally, UAV-assisted MEC is more agile and can better support on-demand computing services than the traditional BS-based MEC.

\subsection{Related Work}
An increasing amount of research attention has been paid to various issues in UAV-assisted MEC networks. There are mainly two categories according to the number of objectives to optimize, namely single- and multi-objective optimization.

\subsubsection{Single-Objective Optimization}
There has been a large amount of research studying single-objective optimization (SOO) problems in the context of UAV-assisted MEC, where only one objective is considered for optimization, e.g., delay or enery consumption. Traditional methods and deep reinforcement learning (DRL) are mainstream optimization techniques.

\textbf{SOO with traditional methods}. Liu \textit{et al}. \cite{liu2019uav} investigated the computation offloading and UAV trajectory planning problem, with the total energy consumption of UAVs minimized. The authors used a convex optimization method to address it. Zhang \textit{et al}. \cite{zhang2019joint} emphasized task offloading and UAV relay communication in an MEC system with one UAV and one BS, where the successive convex approximation technology was adopted to minimize the system's energy consumption. The same technology was also used in \cite{sun2021joint} to reduce the energy consumption of a UAV by optimizing its trajectory and offloading schedule. Tun \textit{et al}. \cite{tun2020energy} proposed a successive convex method that minimized the energy consumption of IoT devices and UAVs, with the task offloading decision and UAVs' trajectories taken into account. Apostolopoulos \textit{et al}. \cite{apostolopoulos2021data} presented a data offloading decision-making framework consisting of ground and UAV-assisted MEC servers and the authors applied convex optimization to maximize each user’s satisfaction utility. Ye \textit{et al}. \cite{ye2020offspeeding} studied the energy-efficient flight speed scheduling problem, with the purpose of minimizing the UAV's energy consumption. The authors obtained near-optimal solutions to UAV's flight speed scheduling via heuristics. In \cite{zhang2018stochastic}, a Lyapunov-based method was developed to minimize the average energy consumption of UAVs, where the task offloading and UAV trajectory were taken into account.

\textbf{SOO with DRL methods}. Zhou \textit{et al}. \cite{zhou2020deep} proposed a deep risk-sensitive reinforcement learning (RL) algorithm to minimize the total delay of all tasks while satisfying UAV's energy capacity constraint. Chen \textit{et al}. \cite{chen2020age} developed a DRL-based online method to maximize the long-term computation performance, where two deep Q-networks (DQN) were adopted. Zhao \textit{et al}. \cite{zhao2020deep} studied the UAV trajectory planning and power allocation problem and applied deep deterministic policy gradient (DDPG) to maximize the long-term network utility. Based on double deep Q-network (DQN), Liu \textit{et al}. \cite{liu2020cooperative} proposed a two-phase DRL offloading algorithm for multi-UAV systems, with the system's total utility maximized. To minimize the total resource consumption of SDs, Wang \textit{et al}. \cite{wang2020intelligent} presented an intelligent resource allocation method based on multi-agent Q-learning. In \cite{ren2021enabling}, a hierarchical DRL algorithm was developed to minimize the average delay of tasks by jointly optimizing the movement locations of SDs and offloading decisions. To minimize the energy consumption of all SDs, Wang \textit{et al}. \cite{wang2021deep} presented a trajectory control method based on DDPG with prioritized experience replay. Dai \textit{et al}. \cite{dai2021towards} considered a UAV-and-BS enabled MEC system and devised a DDPG-based task association scheduling method to minimize the system's energy consumption.

\subsubsection{Multi-Objective Optimization}
In nature, multiple possibly conflicting objectives exist in UAV-assisted MEC. For example, one should consider the tradeoff between delay and energy consumption in the task offloading decision-making process; one should balance the energy consumption and flying speed when planning a UAV's trajectory. Some research efforts have been dedicated to multi-objective optimization (MOO) problems. 

\textbf{MOO with traditional methods}. In \cite{zhang2020energy}, a game-theory-based method was proposed to optimize the weighted cost of delay and energy consumption in UAV-assisted MEC with multiple SDs and single UAV, subject to the resource competition constraint. Ning \textit{et al}. \cite{ning2021dynamic} considered the computation offloading and server deployment problem and designed two stochastic game methods to minimize the computation delay and energy consumption of each UAV. Zhan \textit{et al}. \cite{zhan2020completion} studied the computation offloading and resource allocation problem and designed a successive convex optimization method to minimize the energy consumption and completion delay of a UAV. Lin \textit{et al}. \cite{lin2022novel} developed a Lyapunov based resource allocation method for UAV-assisted MEC systems, aiming at reducing the overall energy consumption and computation delay.

\textbf{MOO with DRL methods}. In \cite{yang2020multi}, to improve the task execution efficiency of each UAV, a DQN-based task scheduling algorithm was proposed to balance between the network load and task execution delay. Chen \textit{et al}. \cite{chen2020resource} considered a three-dimensional UAV-assisted MEC system, minimizing the task processing delay and energy consumption by double DQN. In \cite{zhang2021task}, DQN was used to minimize the energy consumption and computation delay of MEC networks simultaneously. Sun \textit{et al}. \cite{sun2021aoi} studied a bi-objective optimization problem with the age-of-information (AoI) and UAV's energy-consumption as two objectives to minimize and devised a twin-delayed DDPG (TD3) for UAV trajectory control. Wang \textit{et al}. \cite{wang2020multi} proposed a multi-agent DDPG based trajectory control algorithm that took the geographical fairness among UAVs and energy consumption of SDs as two objectives for optimization. Peng \textit{et al}. \cite{peng2021deep} studied the single-UAV trajectory control problem and adopted double DQN to minimize the UAV's energy consumption and maximize the amount of offloaded data, simultaneously.

\subsubsection{Analysis and Motivation}
Despite the ample research efforts dedicated, UAV-assisted MEC still faces great challenges in terms of system design and optimization. We discuss these challenges from two aspects, i.e., system modeling and optimization techniques.

\textbf{System modeling}. In most existing works, see \cite{zhan2020completion, ning2021dynamic, yang2020multi, wang2020multi}, a system only employs one or more UAVs for task collection and processing, where no base station is involved. Although it suffices in cases where the number of SDs is small, such a system cannot support large-scale MEC deployment since UAVs usually have limited computing resources. Multiple UAVs could alleviate the computing pressure at an increased deployment cost. To handle the issue, some works\cite{zhang2019joint, apostolopoulos2021data} focus on UAV-assisted MEC systems, where BSs are considered. With efficient collaboration between UAV and BS, various computing services can be provisioned to multiple SDs. UAV-assisted MEC involving BSs is a practical scenario. 

In some extreme scenarios, SDs cannot be reached by BS due to natural disasters, military attacks or simply being out of BS's coverage. In this case, a UAV has two roles to play: (1) an MEC server that runs some of the collected computation tasks from SDs and sends back results to them, or (2) a relay that forwards some computation tasks to a BS. However, This scenario has received little research attention in the literature. That is our motivation to consider a UAV-assisted MEC system without direct connection between SDs and BSs. 

On the other hand, considering delay and energy consumption as optimization objectives is one of the main research streams on UAV-assisted MEC. Most existing works optimize the two individually. The fact that the conflicts between objectives are neglected easily leads to biased optimization results. Meanwhile, a few studies focus on the maximization of the number of tasks collected by UAV(s), which also reflects the benefits that an MEC system brings to us. Therefore, delay, energy consumption and number of tasks collected are three important concerns when designing UAV-assisted MEC systems. However, to the best of our knowledge, little research has been dedicated to a system with these three objectives taken into account. That is why we are motivated to emphasize the UAV-assisted MEC system with delay, energy consumption and number of tasks collected as three objectives for optimization.

\textbf{Optimization techniques.} Traditional methods, including convex optimization \cite{liu2019uav, zhang2019joint, sun2021joint, tun2020energy, apostolopoulos2021data, zhan2020completion}, heuristics \cite{ye2020offspeeding}, Lyapunov optimization \cite{zhang2018stochastic, lin2022novel}, and game theory \cite{zhang2020energy, ning2021dynamic}, work well when dealing with various optimization issues under static scenarios, such as a UAV hovering over a fixed spot during the whole flying mission. However, these methods are hardly adapted to a dynamic environment, especially when UAVs move quickly and tasks arrive unpredictably. That is because the dynamics and uncertainty will frequently trigger execution of the above methods that launch from scratch, resulting in high computational burdens and slow response. Thus, these methods are not suitable for always responding quickly to users while the MEC environment is ever-changing.

Different from the traditional methods, DRL can deal with complicated control problems with little prior information extracted from dynamic MEC scenarios. The reason is that DRL methods are able to quickly adapt their behaviors to the changes by interacting with the corresponding environment. However, all the DRLs above are single-objective RL (SORL), which defines the user utility as a linear scalarization based on preferences (i.e., weights) across objectives. These SORL methods first aggregate multiple objectives into a scalar reward via weighted sum and then optimize the reward. Nevertheless, the conflicts between objectives are ignored because weighted sum is usually biased and hardly strikes a balance between objectives.

Multi-objective RL (MORL) can well address the challenge above \cite{abels2019dynamic, xu2020prediction}. According to the number of learned policies, MORLs can be divided into two categories, namely single-policy MORLs and multi-policy MORLs. A single-policy MORL aims to optimize one policy for a given preference. For example, the authors in \cite{yu2021multi} extended a single-objective DDPG to a single-policy MORL to optimize the data rate, total harvested energy, and UAV's energy consumption. However, a single-policy MORL cannot output multiple optimal policies after a run, each of which optimizes a certain preference.

Unlike single-policy MORLs, multi-policy MORLs can learn a set of policies that approximate the true Pareto front. These policies correspond to different tradeoffs, and the decision maker can select the one that matches with the current preference. With the multi-task multi-objective proximal policy optimization (PPO), the evolutionary MORL (EMORL) algorithm \cite{xu2020prediction} has promising potential to find a set of high-quality policies. This algorithm has been successfully applied to continuous robotic control problems. This is why we adapt EMORL to the UAV-assisted MEC concerned in this paper.

\subsection{Contribution}
This paper studies the trajectory control and task offloading (TCTO) problem in a UAV-assisted MEC system, where a UAV and a BS work together to provide SDs with computing services. We consider the scenario that SDs are not directly connected by the BS and the UAV plays as an MEC server when processing a collected computation task locally or a wireless relay when forwarding the task to the BS. The UAV collects computation tasks from the SDs within its coverage and decides the proportion of these tasks to be offloaded to the BS for remote processing. Different from the existing works that either optimize a single objective or a number of objectives via weighted sum, this paper considers three conflicting objectives and aims to optimize them, simultaneously. To obtain a set of Pareto optimal policies, we adapt EMORL to the MOO problem. The main contributions are summarized as follows. 

\begin{itemize}
	\item  We formulate the TCTO problem as an MOO problem, aiming at minimizing the task delay and UAV's energy consumption, and maximizing the number of tasks collected by the UAV, simultaneously. The MOO problem is difficult to address because the three objectives conflict with each other and to strike a balance between them is quite challenging.
	
	\item We model a multi-objective Markov decision process (MOMDP) with a vector reward of three elements for the TCTO problem, where each element corresponds to an optimization objective. Based on the MOMDP model, we adapt EMORL to the TCTO problem, namely EMORL-TCTO. EMORL-TCTO can output multiple policies to satisfy various preferences of users at a run. To our knowledge, this is the first work that applies a multi-policy MORL to the UAV-assisted MEC field.
	
	\item We conduct extensive experiments using six test instances. The results clearly show that the proposed EMORL-TCTO can obtain a set of high-quality nondominated policies and outperforms two state-of-the-art multi-objective evolutionary algorithms and two exclusively devised multi-policy MORLs against several evaluation criteria, including the inverted generational distance, hyper volume, average comprehensive objective indicator, and Friedman test. 
\end{itemize}

The remainder of the paper is organized as follows. The system model and problem formulation are presented in Section \ref{system model}. In Section \ref{background}, we briefly review the MOMDP and MOO. In Section \ref{the proposed algorithm}, we introduce the proposed algorithm for the TCTO problem in detail. Section \ref{simulation results} analyzes and discusses the simulation results. Finally, Section \ref{conclusion} concludes the paper.

\section{System Model and Problem Formulation}
\label{system model}
As shown in Fig. 1, this paper considers a UAV-assisted MEC system consisting of one UAV, one BS, and a set of SDs. These SDs are randomly scattered in a rectangular area and their computation tasks arrive dynamically. A rotary-wing UAV can hover in the air and fly at a low altitude sufficiently close to SDs. Considering the economical and scalable deployment, this paper considers a rotary-wing UAV with limited computing resources. The UAV is responsible for task collection, i.e., it flies along a planned trajectory to collect computation tasks from SDs within its coverage. It either executes all these tasks locally or offloads a proportion of them to the BS for processing when needed. The BS has abundant computing resources and acts as a complementary offloading solution to the UAV. 

\begin{figure}
	\centering
	\includegraphics[scale=0.3]{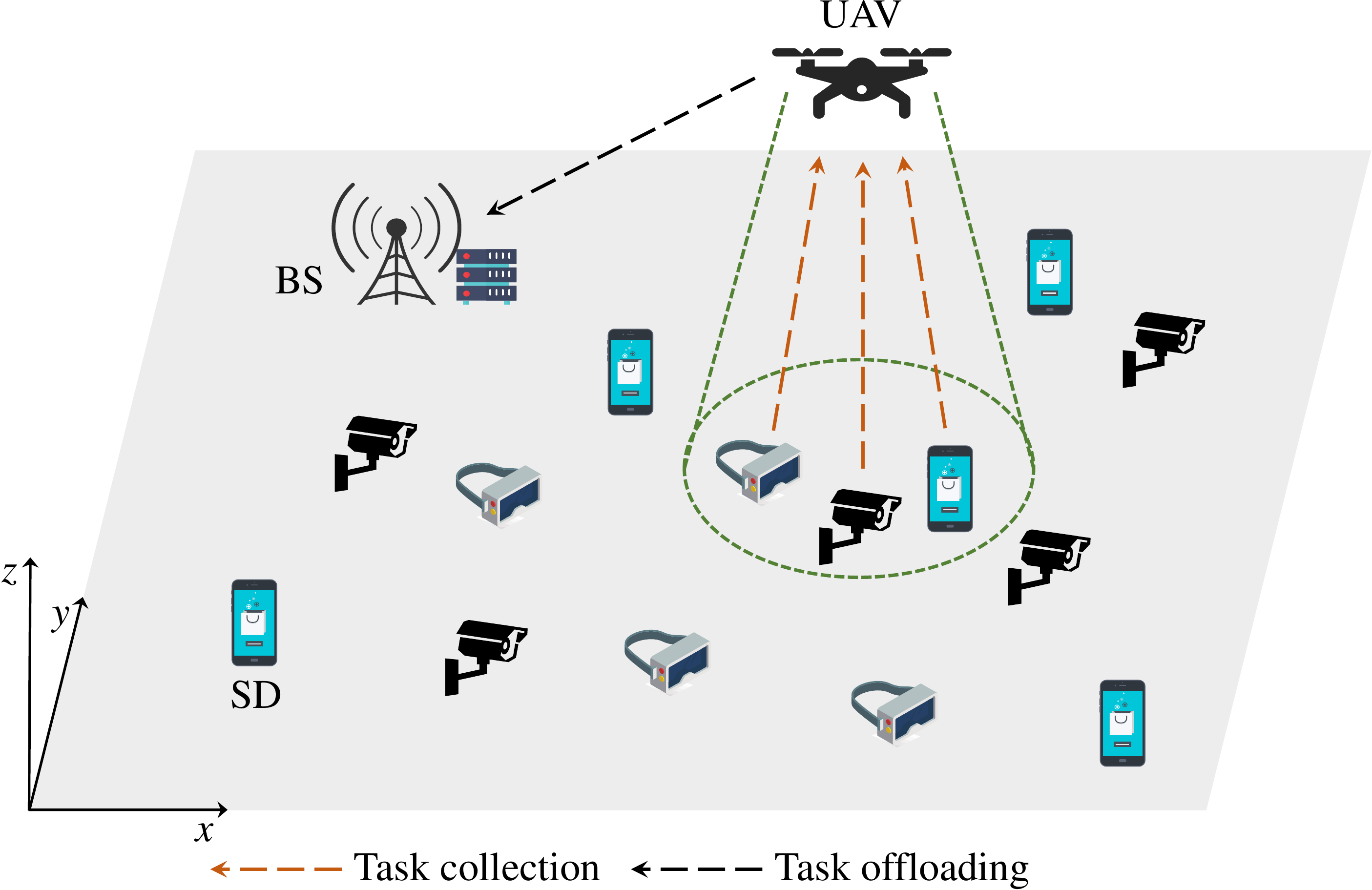}
	\caption{UAV-assisted MEC system.}
	\label{Figure: MEC system}
\end{figure}

We consider a discrete time system, where each time slot has a time duration of $\tau$. Suppose the entire task collection process of the UAV lasts for $T$ time slots. Let $\T=\{1,...,T\}$ denote the set of time slots. Let $\K=\{1,...,K\}$ be the set of SDs, where $K$ is the number of SDs. The main notations used in this paper are summarized in Table \ref{Table: main notations}.

\begin{table}
	
	\centering
	\caption{Summary of main notations}
	\begin{tabular}{ll}  
		\hline  
		Notation & Definition \\  
		\cline{1-2}
		\multicolumn{2}{c}{ Notation used in system model} \\
		\hline
		
		$b_t$ & Offloading decision of the UAV in time slot $t$\\
		$d_{\rm max}$ & Maximal distance the UAV can move in each time slot\\
		$d_t$ & Horizontal distance the UAV flies in time slot $t$\\
		
		$f_{\rm U}$ & Computing capability of the UAV\\ 
		$H$ & Fixed flying altitude of the UAV\\
		
		$k$ & The $k$-th SD \\
		$K$ & Number of SDs \\
		$\K$ & Set of SDs\\ 
		$\K_t^k$ & Set of SDs covered by the UAV in time slot $t$\\
		
		$l_t^k$ & Task arrival indicator of SD $k$ in time slot $t$\\
		$L_{\rm max}$ & Maximum number of tasks allowed to be stored by an SD\\
		$L_t^k$ & Number of tasks in the $k$-th SD's queue in time slot $t$ \\
		
		$N_{\rm max}$ & Maximum number of tasks in the computing queue\\
		$N_t^{\rm c}$ & Number of collected tasks from SDs in time slot $t$\\
		$N_t^{\rm O}$ & Number of tasks offloaded to the BS in time slot $t$\\
		$N_t^{\rm L}$ & Number of tasks executed by the UAV in time slot $t$\\
		$P_{\rm U}$ & Transmission power of the UAV\\
		$R_{\rm max}$ & Maximum horizontal coverage of the UAV\\
		$T$ & Number of time slots \\
		$\T$ & Set of time slots\\
		$W$ & Channel bandwidth\\
		
		$\alpha$ & Input data size of a task\\
		$\beta$  & Number of CPU cycles required to process a task\\
		$\theta_{\rm max}$ & Maximal azimuth angle of the UAV\\
		$\theta_t$ & Horizontal direction the UAV flies in time slot $t$ \\
		$\zeta^k$ & Parameter of Bernoulli random variable of SD $k$\\
		$\kappa$ & Effective capacitance coefficient\\
		$\mu_t$ & Data rate of the wireless channel in time slot $t$\\
		$\sigma^2$ & Background noise power\\
		$\tau$ & Time duration of a time slot\\
		$\phi$ & Number of tasks processed by the UAV within a time\\
		& slot\\
		
		\hline
		\multicolumn{2}{c}{ Notation used in reinforcement learning} \\
		\hline
		$a$ & Action\\
		$\mathcal{A}$ & Action space\\
		$\mathbf{A}_t$& Vector-valued advantage function \\
		$A^{\mathbf{w}_i}_t$ & Extended advantage function with weight vector $\mathbf{w}_i$\\ 
		$\mathbf{F}(\pi)$ & Objective vector of policy $\pi$\\ 
		$n$ & Number of learning tasks\\
		$\textbf{r}_t$ & Vector-valued reward at time step $t$\\
		$\textbf{R}_\pi$ & Vector-valued return following policy $\pi$\\
		$s$ & State\\
		$\mathcal{S}$ & State space\\
		$\textbf{V}_\pi(s)$ & Multi-objective value function in state $s$\\
		$\mathcal{W}$ & Set of evenly distributed weight vectors\\
		$\lambda$ & Parameter of general advantage estimator\\
		$\gamma$ & Discount factor\\
		$\Gamma_i$ & The $i$-th learning task in $\Omega$, $i=1,...,n$\\
		$\Omega$ & Set of learning tasks\\
		
		\hline
	\end{tabular}
\label{Table: main notations}
\end{table}

\subsection{Task Model}
We assume that the computation tasks arriving at SD $k \in \K$ can be modeled as an independent and identically distributed sequence of Bernoulli random variables with parameter $\zeta^k \in [0, 1]$. Different SDs are associated with different parameters of Bernoulli random variables. Let $l^k_t$ denote the task arrival indicator of SD $k$ in time slot $t$. $l^k_t = 1$ if a task is generated at the beginning of $t$ and $l^k_t = 0$, otherwise. We have $ {\rm Pr}(l^k_t = 1) = 1 - {\rm Pr}(l^k_t = 0) = \zeta^k$, where ${\rm Pr}(\cdot)$ stands for the probability of an event occurring. A computation task is modeled as tuple $\langle \alpha, \beta \rangle$, where $\alpha$ denotes the input data size of the task and $\beta$ is the number of CPU cycles required to process the task. For an arbitrary SD, a computation task generated in $t$ is stored in its task queue. Let $L^k_t$ be the number of tasks in the $k$-th SD's queue waiting to be uploaded in $t$, which is updated by 
\begin{equation}
	L^k_{t+1} = \min\{L^k_t + l^k_t, L_{\rm max}      \},
\end{equation}
where $L_{\rm max}$ is the maximum number of tasks allowed to be stored in the $k$-th SD's queue. If the queue is full, each newly arrival task is dropped. Hence, it is of great significance for SDs to upload their computation tasks to the UAV in time. In this paper, the time division multiple access protocol is adopted for uploading computation tasks.

\subsection{UAV Movement Model}
We assume that the UAV flies at an altitude of $H$, where $H$ is a positive constant. Let $\theta_t$ and $d_t$ denote the horizontal direction and distance with which the UAV flies in time slot $t$, respectively, with the following constraints met:
\begin{equation}
	0 \leq \theta_t \leq 2\pi, 0 \leq d_t \leq d_{\rm max},
\end{equation}
where $d_{\rm max}$ represents the maximal flying distance that the UAV can move in each time slot due to the limited power budget.

Similar to previous studies \cite{wang2021deep, yu2021multi}, we adopt the Cartesian coordinate system to model the movement of the UAV. Let $\mathbf{c}_t^{\rm U} = [x_t^{\rm U}, y_t^{\rm U}]$ denote the UAV's horizontal coordinate in time slot $t$. Based on $\theta_t$ and $d_t$, we obtain the UAV's horizontal coordinate in time slot $t+1$ by 
\begin{equation}
	\begin{cases} x_{t+1}^{\rm U}=x_t^{\rm U} + d_t \cdot \cos(\theta_t) \\
		y_{t+1}^{\rm U}=y_t^{\rm U} + d_t \cdot \sin(\theta_t).
	\end{cases}
\end{equation}
Assume that the UAV flies at a constant velocity $v_t = {d_t}/{\tau}$, limited by a pre-defined maximum flying velocity $v_{\rm max}$. The UAV can only move within a rectangular area whose side lengths are $x_{\rm max}$ and $y_{\rm max}$. We have
\begin{equation}
	0 \leq x_t^{\rm U}	\leq x_{\rm max}, 0 \leq y_t^{\rm U}	\leq y_{\rm max}.
\end{equation} 

When a rotary-wing UAV flies, its propulsion power consumption with speed $v$, $P(v)$, is defined as \cite{yu2021multi}
\begin{equation}
	\begin{split}
		P(v) 
		& = P_1 \left( 1+ \frac{3v^2}{U_{\rm tip}^2} \right) + P_2\left(\sqrt{1+ \frac{v^4}{4v_0^4}} - \frac{v^2}{2v_0^2} \right)^{1/2} \\
		& + \frac{1}{2} d_0 \rho g A v^3.
	\end{split}
	\label{Eq: UAV speed}
\end{equation}

It is seen that $P(v)$ consists of three parts: the blade profile, induced power, and parasite power. $P_1$ and $U_{\rm tip}$ denote the blade profile power under hovering status and tip speed of rotor blade, respectively. $P_2$ and $v_0$ represent the induced power and mean rotor induced velocity in hovering, respectively. As for the parasite power, $d_0$, $\rho$, $g$, and $A$ indicate the fuselage drag ratio, air density, rotor solidity, and rotor disc area, respectively. Note that when the UAV hovers (i.e., $v=0$), the corresponding power consumption $P_{\rm h}$ is the summation of $P_1$ and $P_2$. The energy consumption when the UAV is flying and hovering during a time duration of $T$, $E_{\rm fh}$, is obtained by
\begin{equation}
	E_{\rm fh} = \int_0^T P(v_t) dt.
	\label{Eq: energy of flying and hovering}
\end{equation}

\subsection{Computing Model}
\subsubsection{Local Computing}
Assume the UAV maintains a computing queue that stores the computation tasks collected from SDs awaiting for further processing. As the UAV can stay at a low altitude sufficiently close to SDs, this paper ignores the delay for collecting the computation tasks in each time slot, so does the corresponding receiving power consumption at the UAV. In this paper, the delay for processing tasks locally on the UAV in time slot $t$ consists of the local processing and queuing delays. Let $N_t^{\rm u} \in [0, N_{\rm max}]$ represent the number of uncompleted tasks in the computing queue at the beginning of $t$, where $N_{\rm max}$ is the maximum number of tasks allowed. Let $b_t \in [0, 1]$ be the proportion of tasks in the computing queue to be offloaded to the BS in $t$, namely the UAV's offloading decision for $t$. Specifically, the UAV offloads $N_t^{\rm O} = \lfloor b_t N_t^{\rm u} \rfloor$ computation tasks to the BS for remote processing, where $\lfloor \cdot \rfloor$ denotes the floor function. The remaining $N_t^{\rm L} = N_t^{\rm u} -  N_t^{\rm O}$ computation tasks are locally executed on the UAV. Let $\phi = \lfloor {\tau f_{\rm U}}/{\beta} \rfloor$ denote the number of computation tasks processed by the UAV within each time slot, where $f_{\rm U}$ denotes the UAV's computing capability. Based on $N_t^{\rm u}$ and $N_t^{\rm O}$, the number of queueing tasks in the computing queue at the end of $t$, $N_t^{\rm q}$, is defined as
\begin{equation}
	N_t^{\rm q} = \max \left\{N_t^{\rm u} - \phi -  N_t^{\rm O}, 0\right\}.
\end{equation}

Let $\mathbf{c}^k = [x^{k}, y^{k}]$ be the horizontal coordinate of SD $k\in \K$. The UAV can only collect the tasks within its coverage area. Let $\K_t^{\rm c}$ represent the set of SDs covered by the UAV in time slot $t$, which is defined as
\begin{equation}
	\K_t^{\rm c} = \{k| d_t^k  \leq  R_{\rm max}, k\in \K  \},
	\label{Eq: collected task num}
\end{equation}
where $d_t^k = \sqrt{(x_t^{\rm U} - x^{k})^2 + (y_t^{\rm U} - y^{k})^2}$ is the horizontal distance between the UAV and SD $k$ in $t$. $R_{\rm max}$ is the UAV's maximal horizontal coverage, given that it has a maximal azimuth angle $\theta_{\rm max}$ \cite{wang2021deep}. $R_{\rm max}$ is calculated by 
\begin{equation}
	R_{\rm max} = H \cdot \tan(\theta_{\rm max}).
	\label{Eq: coverage}
\end{equation}
Based on Eq. (\ref{Eq: collected task num}), the number of tasks collected by the UAV in $t$ is obtained by
\begin{equation}
	N_t^{\rm c} = \sum_{k\in \K_t^{\rm c} } L_t^k.
	\label{Eq: collected task number}
\end{equation}
The number of uncompleted tasks to be processed in $t+1$, $N_{t+1}^{\rm u}$, is updated at the end of $t$ as 
\begin{equation}
	N_{t+1}^{\rm u} = \min\{N_t^{\rm q}+N_t^{\rm c}, N_{\rm max}\}.
\end{equation}

In $t$, the delay for completing the $N_t^{\rm L}$ tasks locally on the UAV can be calculated by 
\begin{equation}
	D_t^{\rm L} = \frac{\min\{\phi, N_t^{\rm L}\} \beta}{f_{\rm U}}   + \tau N_t^{\rm q} .
	\label{Eq: local delay}
\end{equation}
There are two parts in Eq. (\ref{Eq: local delay}). The first part, ${\min\{\phi, N_t^{\rm L}\} \beta}/f_{\rm U}$, is the local processing delay, and the second one, $\tau N_t^{\rm q}$, is the queuing delay of all $N_t^{\rm q}$ tasks waiting in the computing queue. The corresponding energy consumption of the UAV is calculated by
\begin{equation}
	E_t^{\rm L} = \kappa \cdot \min\{\phi, N_t^{\rm L}\} \beta \cdot (f_{\rm U})^2, 
	\label{Eq: local energy}
\end{equation}
where $\kappa$ is the effective capacitance coefficient depending on the chip structure used.

\subsubsection{Task Offloading}
The UAV allows a proportion of its collected tasks to be offloaded to the BS for remote processing. According to the Shannon-Hartley theorem \cite{zhou2020deep}, we define the data rate of the wireless link between the UAV and BS in $t$ as
\begin{equation}
	\mu_t = W \cdot \log_2 \left(1 + \frac{P_{\rm U} \cdot 10^{\frac{PL(d_t^{\rm UB}, \theta_t^{\rm UB})}{10}}}{\sigma^2} \right),
	\label{Eq: data rate}
\end{equation}
where $W$, $P_{\rm U}$, and $\sigma^2$ are the channel bandwidth of the wireless link, transmission power of the UAV, and background noise power, respectively. Referring to \cite{zhou2020deep}, this paper defines the pathloss between the UAV and BS in $t$ as 
\begin{equation}
	PL(d_t^{\rm UB}, \theta_t^{\rm UB}) = 10A_0 \log(d_t^{\rm UB}) + B_0(\theta_t^{\rm UB} - \theta_0) {\rm e}^{\frac{\theta_0 - \theta_t^{\rm UB}}{C_0}} + \eta_0,
\label{pathloss}
\end{equation}
where $d_t^{\rm UB}$ and $\theta_t^{\rm UB}$ are the distance and vertical angle between the UAV and BS in $t$, respectively. $d_t^{\rm UB}$ and $\theta_t^{\rm UB}$ in Eq. (\ref{pathloss}) are obtained based on the horizontal coordinates of the UAV and BS.

The UAV needs to complete the transmission process of the $N_t^{\rm O}$ computation tasks before it flies out of the BS's coverage. Thus, the time duration $\varphi_t$ that the UAV has been staying in the coverage of the BS since the beginning of $t$ is written as 
\begin{equation}
	\varphi_t = \mathop{\arg\min}_{l} \left(  \sum_{i=t}^{t+l} \tau \mu_i   \geq  \alpha N_t^{\rm O}  \right),
\end{equation}
where $\alpha$ stands for the input data size of a computation task. Let $D_t^{\rm O}$ denote the delay for offloading the $N_t^{\rm O}$ computation tasks to the BS, which is calculated by
\begin{equation}
	D_t^{\rm O} =
	\begin{cases}
		 (\varphi_t-1)\tau  + \frac{\alpha N_t^{\rm O} - \sum_{i=t}^{\varphi_t-1} \tau \mu_i }{ \mu_{t+\varphi_t}}, & \text{if} \ \  \alpha N_t^{\rm O} <  \sum_{i=t}^{\varphi_t}  \tau \mu_i \\
		 \tau \varphi_t,  & \text{if} \ \  \alpha N_t^{\rm O} =  \sum_{i=t}^{\varphi_t}  \tau \mu_i
	\end{cases}
	\label{Eq: offloading delay}
\end{equation} 
The corresponding energy consumption of the UAV is calculated as
\begin{equation}
	E_t^{\rm O} = P_{\rm U} \cdot D_t^{\rm O}.
	\label{Eq: offloading energy}
\end{equation}
Assume that the BS is of rich computing resources. Thus, the delay for processing the tasks on the BS can be neglected. Further, the delay for returning the task results to an SD is also ignored because the computation result of a task is usually much smaller than its input data size.

\subsection{Problem Formulation}
Based on Eqs. (\ref{Eq: local delay}) and (\ref{Eq: offloading delay}), the delay for completing the $N_t^{\rm L}+N_t^{\rm O}$ computation tasks in the UAV's computing queue in $t$ is written as
\begin{equation}
	D_t = D_t^{\rm L} + D_t^{\rm O}.
\end{equation}
Similarly, based on Eqs. (\ref{Eq: local energy}) and (\ref{Eq: offloading energy}), the UAV's energy consumption for local computing and transmitting tasks to the BS in $t$ is defined as
\begin{equation}
	E_t = E_t^{\rm L} + E_t^{\rm O}.
\end{equation}

The total delay for completing all the collected tasks, $D_{\rm total}$, and total energy consumption of the UAV, $E_{\rm total}$, during $T$ time slots are calculated as
\begin{equation}
	D_{\rm total} = \sum_{t = 1}^T D_t,
\end{equation}

\begin{equation}
	E_{\rm total} = \sum_{t = 1}^T E_t + E_{\rm fh}.
\end{equation}
Based on the number of collected tasks defined in Eq. (\ref{Eq: collected task number}) in each time slot, the total number of collected tasks during time duration $T$ can be obtained by
\begin{equation}
	N_{\rm total} = \sum_{t = 1}^{T} N_t^{\rm c}.
\end{equation}

In this work, we aim to minimize the total task delay $D_{\rm total}$ and total energy consumption $E_{\rm total}$, and maximize the total number of tasks collected $N_{\rm total}$, simultaneously, through optimizing the UAV's flying trajectory (i.e., $\theta_t$ and $d_t$) and task offloading decision (i.e., $b_t$), namely the TCTO problem. This problem is an MOO problem in nature, defined as:
\begin{equation}
	\max_{\theta_t, d_t, b_t} (-D_{\rm total}, -E_{\rm total}, N_{\rm total})
\end{equation}

subject to:
\begin{align}
	& \text{C1}:  0 \leq \theta_t \leq 2\pi,   &\quad \forall t\in \T, \nonumber \\
	& \text{C2}:  0 \leq d_t \leq d_{\rm max}, &\quad \forall t\in \T, \nonumber \\
	& \text{C3}:  b_t \in [0, 1], &\quad \forall t\in \T, \nonumber \\
	& \text{C4}: 0\leq x_t^{\rm U} \leq x_{\rm max},     &\forall  t\in \T,   \nonumber\\
	& \text{C5}: 0\leq y_t^{\rm U} \leq y_{\rm max},     &\forall  t\in \T,   \nonumber\\
	& \text{C6}: d_t^k \leq R_{\rm max},     &\forall  k \in \K_t^{\rm c}, t\in \T.   \nonumber
\end{align}
Constraints C1 and C2 confine the horizontal direction and distance of a flying UAV. Constraint C3 specifies that the offloading decision for time slot $t$ is a variable between 0 and 1. Constraints C4 and C5 together specify the UAV's movement area. Constraint C6 ensures that the UAV can only collect computation tasks from SDs within its coverage. 

It is easily understood that to increase $N_{\rm total}$, the UAV should fly with an appropriate trajectory so that it can cover as many SDs and collect their computation tasks as possible. However, the more the computation tasks collected, the higher the energy consumption incurred on the UAV because more tasks need to be handled by the UAV. Admittedly, offloading helps to reduce the UAV’s energy consumption as some tasks are processed by the BS. However, it results in additional transmission delays. So, one can easily observe that the three objectives, i.e., minimization of $D_{\rm total}$, minimization of $E_{\rm total}$, and maximization of $N_{\rm total}$, conflict with each other. Unfortunately, traditional SORLs cannot optimize every objective in a single run since these methods all aggregate multiple objectives into one via weighted sum, nor can they change the weights across objectives during the run, being not easy to strike a balance between them. On the other hand, single-policy MORLs only obtain an optimal policy for a pre-defined set of weights after a run. Although these methods are adaptive to the changes of weights for objectives, they cannot output multiple optimal policies in a run, of which each optimizes a certain set of weights. That is why we are motivated to adapt EMROL, an emerging multi-policy MORL, to the TCTO problem concerned in this paper.

\section{Overview of MOMDP and MOO}
\label{background}

This section first recalls the multi-objective Markov decision process (MOMDP). Then, we introduce the multi-objective optimization (MOO) problem.
\subsection{MOMDP}
The TCTO problem is a multi-objective control problem that can be modeled by MOMDP \cite{xu2020prediction}. An MOMDP is defined by tuple $\langle \mathcal{S}, \mathcal{A}, \mathcal{Q}, \mathbf{r}, \gamma, \mathcal{D} \rangle$, where $\mathcal{S}$ is the state space. $\mathcal{A}$ is the action space and $\mathcal{Q}(s^\prime|s,a)$ is the state transition probability. $\mathbf{r}=(r^1,...,r^m)$ is the vector-valued reward function and $m$ is the number of objectives. $\gamma \in [0,1]$ is the discount factor, and $\mathcal{D}$ is the initial state distribution.

In MOMDPs, a policy $\pi: \mathcal{S} \to \mathcal{A}$ is a state-to-action mapping associated with a vector of expected return $\mathbf{R}_\pi = (R_\pi^1,...,R_\pi^m)$, where $R_\pi^j$ is the expected return corresponding to the $j$-th objective, defined as
\begin{equation}
	R_\pi^j = \mathbb{E}_\pi \left[ \sum_{t=1}^T \gamma^{t-1} r^j(s_t, a_t) |s_1 \backsim \mathcal{D}, a_t \backsim \pi(s_t) \right].
\end{equation}
For the TCTO problem, we have $m=3$, namely, $R_\pi^1, R_\pi^2$ and $R_\pi^3$ are associated with $-D_{\rm total}$, $-E_{\rm total}$, and $N_{\rm total}$, respectively.

The value function $\mathbf{V}_\pi(s): \mathcal{S} \to \mathbb{R}^m$ maps a state $s$ to the vector of expected return under policy $\pi$, defined as
\begin{equation}
	\mathbf{V}_\pi (s) = \mathbb{E}_\pi \left[ \sum_{k=t}^T \gamma^{k-t} \mathbf{r}_k | s_t = s\right],
\end{equation}
where $\mathbf{r}_k=(r_k^1,...,r_k^m)$ denotes the immediate vector-valued reward at time step $k$. Because each element of $\textbf{r}_k$ corresponds to a particular objective, $\mathbf{V}_\pi (s)$ is a multi-objective value function.

\subsection{MOO}
An MOO problem \cite{xu2020prediction} can be formulated as
\begin{align}
	&\max_\pi \mathbf{F}(\pi) = \max_\pi (f^1(\pi),...,f^m(\pi)), \nonumber\\
	&\text{subject to:}\quad \pi \in \Pi.
\end{align}
where $\pi$ is a policy in search space $\Pi$. In objective vector $\mathbf{F}(\pi)$, there are $m$ objective functions, and they generally conflict with each other. Note that the objective value $f^j(\pi)$ is set to $R_\pi^j$, $j=1,...,m$.

Let $\pi_1, \pi_2 \in \Pi$ denote two different policies. $\pi_1$ is said to dominate $\pi_2$, denoted by $\pi_1 \succ \pi_2$, if and only if $f^j(\pi_1) \geq f^j(\pi_2)$ for all $j =1,...,m$, and $f^l(\pi_1)>f^l(\pi_2)$ for at least one index $l \in \{1,...,m\}$. A policy $\pi^\ast \in \Pi$ is Pareto optimal if it is not dominated by any other policies in $\Pi$. All Pareto optimal policies form a Pareto optimal set whose mapping in the objective space is known as the Pareto front.

\section{EMORL-TCTO for Trajectory Control and Task Offloading}
\label{the proposed algorithm}
This section first introduces the MOMDP model for the TCTO problem and then describes the proposed EMORL-TCTO algorithm in detail.

\subsection{MOMDP Model}

To address the TCTO problem by an MORL, we need an MOMDP model for the problem first.  The state space, action space, and reward function are described one by one. 

\subsubsection{State Space}
\begin{equation}
	\mathcal{S} = \{s_t|s_t = (\mathbf{c}_t^{\rm U}, N_t^{\rm u}, N_t^{\rm c}), \forall t \in \T    \},
\end{equation}
where $\mathbf{c}_t^{\rm U} = [x_t^{\rm U}, y_t^{\rm U}]$ is the horizontal coordinate of the UAV in time slot $t$. $N_t^{\rm u}$ is the number of uncompleted tasks at the beginning of $t$, and $N_t^{\rm c}$ is the number of newly collected tasks from SDs in $t$.

\subsubsection{Action Space}
\begin{equation}
	\mathcal{A} = \{a_t | a_t = (\theta_t, d_t, b_t), \forall t \in \T    \},
\end{equation}
where $\theta_t$ and $d_t$ denote the horizontal direction and distance with which the UAV flies in $t$, respectively, and $b_t$ is the UAV's offloading decision in $t$.

\subsubsection{ Reward Function}
\begin{equation}
	\mathbf{r}_t = (r_t^{\rm D}, r_t^{\rm E}, r_t^{\rm N})=
	\begin{cases}
		(-D_t, -\frac{E_t}{100}, N_t^{\rm c}), \quad  &\text{if} \ \ \mathds{1}_t=1\\
		(-4D_t, -\frac{E_t}{25}, -2N_t^{\rm c}), & \text{otherwise}
	\end{cases}
\end{equation}
where $D_t$, $E_t$, and $N_t^{\rm c}$ are the delay, energy consumption, and number of tasks collected in time slot $t$, respectively. $\mathds{1}_t$ is an indicator variable that equals 0 if the UAV flies out of the rectangular area in $t$ and $\mathds{1}_t$ is equal to 1, otherwise.

Based on the vector-valued reward $\mathbf{r}_t$, we obtain the return which is the summation of the discounted reward generated at each time step over the long run. Let $\mathbf{R}_\pi = (R_\pi^{\rm D}, R_\pi^{\rm E}, R_\pi^{\rm N})$ be the return of $r_1^{\rm D}$, $r_1^{\rm E}$, and $r_1^{\rm N}$ under policy $\pi$ at the first time step, defined as
\begin{align}
	R_\pi^{\rm D} 
	&= -\sum_{t=1}^T \gamma^{t-1} (\mathds{1}_t  D_t  + 4(1-\mathds{1}_t) D_t) \nonumber \\
	& = -\sum_{t=1}^T \gamma^{t-1} (4-3\mathds{1}_t) D_t,  
\end{align}
\begin{align}
	R_\pi^{\rm E} 
	&= -\sum_{t=1}^T \gamma^{t-1} \left(\frac{\mathds{1}_t}{100}  E_t  + \frac{1-\mathds{1}_t}{25}  E_t \right) \nonumber \\
	& = -\sum_{t=1}^T \gamma^{t-1} \left(\frac{4-3\mathds{1}_t}{100} \right) E_t,  
\end{align}
\begin{align}
	R_\pi^{\rm N} 
	&= \sum_{t=1}^T \gamma^{t-1} (\mathds{1}_t  N_t^{\rm c} - 2(1-\mathds{1}_t) N_t^{\rm c}) \nonumber \\
	& = \sum_{t=1}^T \gamma^{t-1} (3\mathds{1}_t - 2) N_t^{\rm c}. 
\end{align}
Maximizing the expected return $\mathbb{E}[\mathbf{R}_\pi]$ is equivalent to minimizing $D_{\rm total}$ and $E_{\rm total}$, and maximizing $N_{\rm total}$, simultaneously.
\begin{figure*}
	\centering
	\includegraphics[scale=0.52]{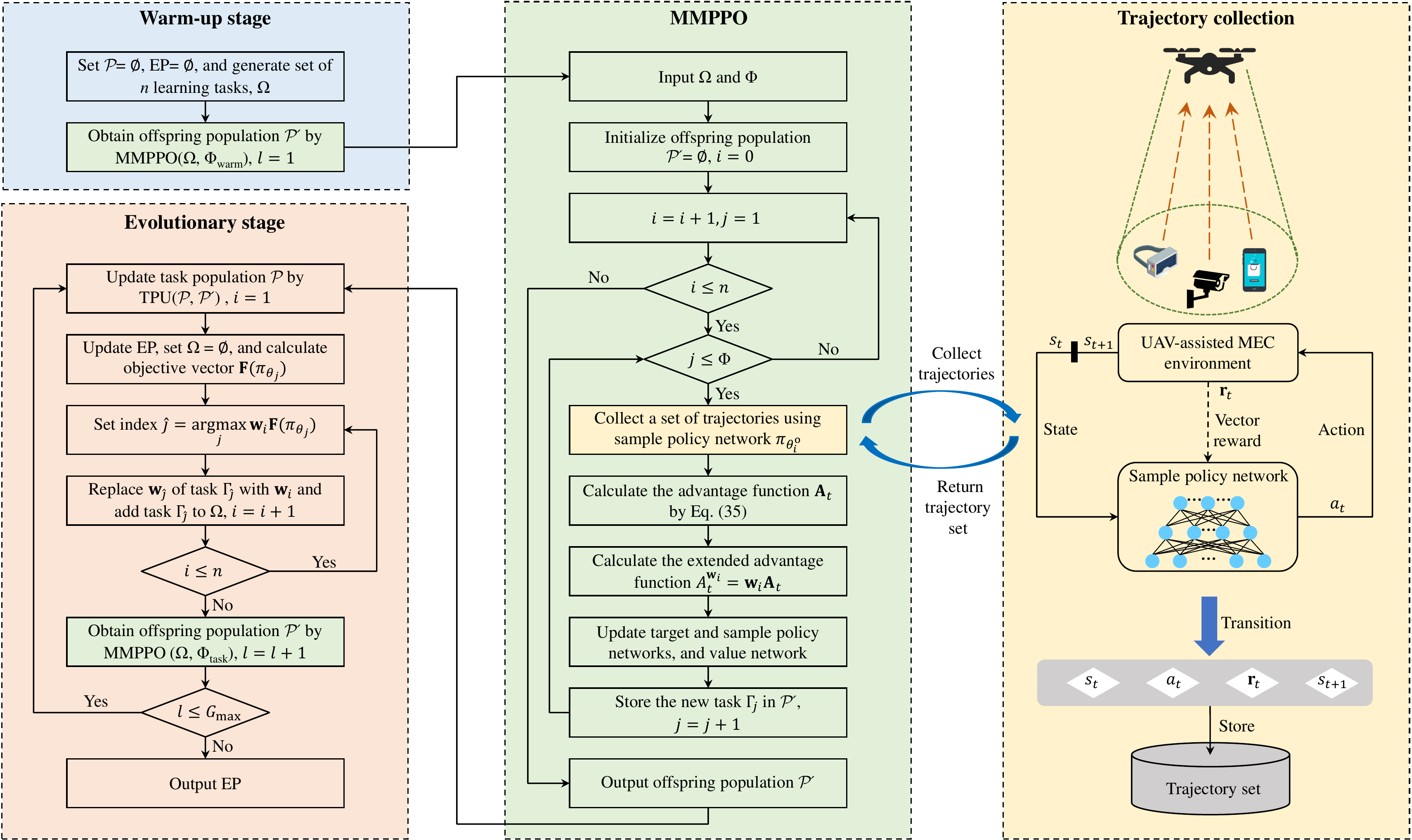} 
	\caption{Framework of EMORL-TCTO.}
	\label{Figure: Framework of EMORL-TCTO}
\end{figure*}

\subsection{EMORL-TCTO Algorithm}
This paper represents a learning task by tuple $\Gamma=\langle \mathbf{w}, \pi_{\theta}, \pi_{\theta^{\rm o}}, \mathbf{V}_{\pi_\theta} \rangle$, where $\mathbf{w} (\sum_{j=1}^m w^{j}=1)$ is the weight vector. $\pi_{\theta}$ is the target policy used to select actions and $\pi_{\theta^{\rm o}}$ is the sample policy used to collect trajectories \footnote{Note that the term ''trajectories'' refers to a sequence of transitions in RL, each of which consists of state, action, reward, and next state. However, the term ''trajectory'' used in the system model represents the UAV's flying path.}. $\mathbf{V}_{\pi_\theta}$ is the multi-objective value function for evaluating the selected actions. Through interacting with the environment, the sample policy $\pi_{\theta^{\rm o}}$ is used to generate the set of trajectories. The generated set is used to update the target policy $\pi_{\theta}$ for several epochs. To avoid a large update of the target policy, a clipped surrogate objective is adopted, which is defined as
\begin{align}
	&J_{\Gamma}^{\rm C}(\theta, \mathbf{w}) = \nonumber \\
	&\mathbb{E} \left[\sum_{t=1}^T  \min\left(\frac{\pi_{\theta} (a_t|s_t)}{\pi_{\theta^{\rm o}} (a_t|s_t)}  A^{\mathbf{w}}_t, {\rm clip}_{1-\epsilon}^{1+\epsilon} \left(\frac{\pi_{\theta} (a_t|s_t)}{\pi_{\theta^{\rm o}} (a_t|s_t)} \right) A^{\mathbf{w}}_t    \right)  \right],
	\label{Eq: surr}
\end{align}
where $A^{\mathbf{w}}_t = \mathbf{w} \mathbf{A}_t$ is the extended advantage function at time step $t$, i.e., the weighted-sum of all elements in the vector-valued advantage function $\mathbf{A}_t$. $\mathbf{A}_t$ is obtained by the general advantage estimator (GAE) \cite{schulman2015high}, defined as
\begin{equation}
	\mathbf{A}_t = \sum_{k=0}^{T-t+1} (\gamma \lambda)^k (\mathbf{r}_{t+k}+\gamma \mathbf{V}_{\pi_\theta} (s_{t+k+1})-\mathbf{V}_{\pi_\theta} (s_{t+k})),
	\label{Eq: GAE}
\end{equation}
where $\lambda \in [0,1]$ is a parameter for tuning the trade-off between variance and bias. ${\rm clip}_{1-\epsilon}^{1+\epsilon}(\Delta)$ is the clip function that constrains the value of $\Delta$, removing the incentive for moving $\Delta$ outside of the interval $[1-\epsilon, 1+\epsilon]$. 

The value function loss is defined as
\begin{equation}
	J_{\Gamma}^{\rm V} (\theta) = \mathbb{E} \left[\sum_{t=1}^T \Vert \mathbf{V}_{\pi_\theta}(s_t) -\mathbf{\widehat V}_{\pi_\theta}(s_{t}) \Vert^2 \right],
	\label{Eq: vaule fun}
\end{equation}
where $\mathbf{\widehat V}_{\pi_\theta}(s_{t}) = \mathbf{r}_t + \gamma \mathbf{V}_{\pi_\theta}(s_{t+1})$ is the target value function.

\begin{figure}
	\centering
	\includegraphics[scale=0.5]{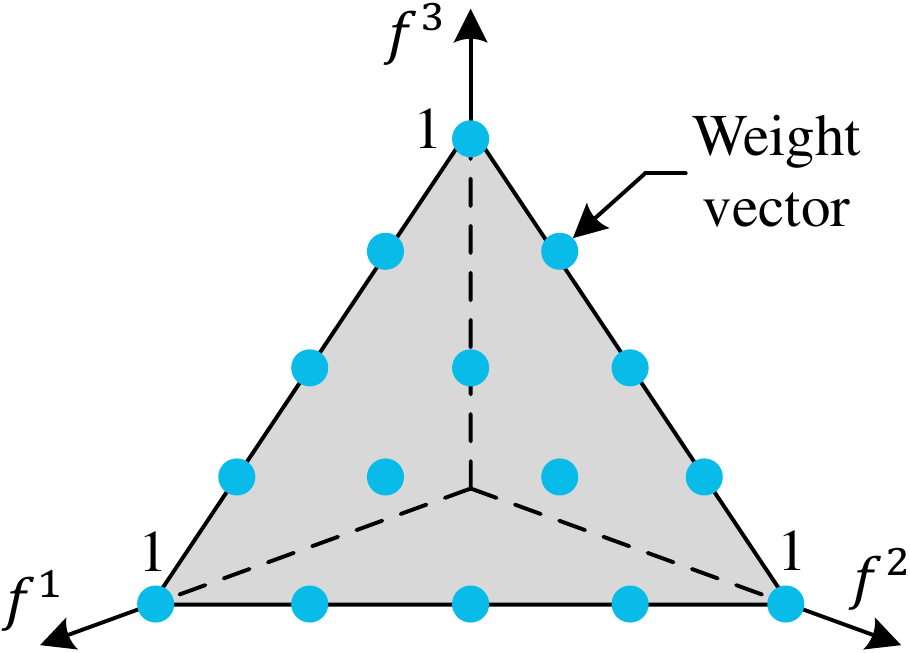}
	\caption{Fifteen evenly distributed weight vectors for a three-objective problem with $\delta=4$.}
	\label{Figure: EDWV}
\end{figure}

The proposed EMORL-TCTO aims to learn a set of Pareto optimal policies through interacting with the environment and its framework is shown in Fig. \ref{Figure: Framework of EMORL-TCTO}. EMORL-TCTO shares the same algorithm structure with the original EMORL \cite{xu2020prediction}. EMORL-TCTO starts from the warm-up stage, where $n$ learning tasks are randomly generated. The offspring population is produced by executing the multi-task multi-objective PPO (MMPPO). Note that each learning task uses its associated sample policy to collect a set of trajectories by interacting with the UAV-assisted MEC environment. After the warm-up stage, EMORL-TCTO proceeds with the evolutionary stage. Both the task population and external Pareto (EP) archive are updated based on the offspring population. Then, we select $n$ new learning tasks from the task population for each weight vector. These tasks are optimized by MMPPO to generate a new generation of the offspring population. The evolutionary stage terminates when a predefined number of generations are completed.

\begin{algorithm}[htb]
	\small
	\caption{Evolutionary multi-objective  reinforcement learning for TCTO problem (EMORL-TCTO)} 
	\begin{algorithmic}[1]
		\REQUIRE
		number of learning tasks $n$, number of warm-up iterations $\Phi_{\rm warm}$, number of task iterations $\Phi_{\rm task}$, number of maximum evolution generations $G_{\rm max}$.\\
		// \textit{Warm-up stage} 
		\STATE Initialize task population $\mathcal{P}=\emptyset$ and external Pareto archive $\rm{EP}=\emptyset$;\\
		\STATE Generate $n$ evenly distributed weight vectors $\{\mathbf w_1,..., \mathbf w_n\}$;
		\STATE Initialize $n$ target policy networks $\{\pi_{\theta_1},...,\pi_{\theta_n}\}$;
		\STATE Initialize the $i$-th sample policy network, $\pi_{\theta_i^{\rm o}} \leftarrow \pi_{\theta_i}, i=1,...,n$;
		\STATE initialize $n$ value networks $\{\mathbf{V}_{\pi_{\theta_1}},...,\mathbf{V}_{\pi_{\theta_n}}\}$;
		\STATE Denote the task set by $\Omega = \{\Gamma_1,...,\Gamma_n\}$, $\Gamma_i= \langle \mathbf{w}_i, \pi_{\theta_i}, \pi_{\theta_i^{\rm o}}, \mathbf{V}_{\pi_{\theta_i}} \rangle$;
		\STATE Obtain offspring population $\mathcal{P}^\prime$ by ${\rm MMPPO}( \Omega, \Phi_{\rm warm})$;
		
		// \textit{Evolutionary stage} 
		\FOR{$ l = 1,..., G_{\rm max}$}
		\STATE Update task population $\mathcal{P}$ by ${\rm TPU}(\mathcal{P}, \mathcal{P}^\prime)$;
		\STATE Update EP based on $\mathcal{P}^\prime$;
		\STATE Set $\Omega = \emptyset$;
		\STATE Calculate $\mathbf{F}(\pi_{\theta_j})$ of target policy $\pi_{\theta_j}$ of each task $\Gamma_j \in \mathcal{P}$;
		\FOR{$\mathbf{w}_i \in \{\mathbf w_1,..., \mathbf w_n\}$} 
		\STATE Set index $\hat j=\arg\max_{j=1,...,|\mathcal{P}|} \{\mathbf{w}_i \mathbf{F}(\pi_{\theta_j})\}$;
		\STATE Replace weight vector $\mathbf{w}_{\hat j}$ of task $\Gamma_{\hat j}$ with $\mathbf{w}_i$;
		\STATE Add task $\Gamma_{\hat j}$ to $\Omega$;
		\ENDFOR
		
		\STATE Obtain offspring population $\mathcal{P}^\prime$ by ${\rm MMPPO}( \Omega, \Phi_{\rm task})$;
		
		\ENDFOR
		
		\ENSURE external Pareto archive $\rm{EP}$.
	\end{algorithmic}
	\label{Algorithm: EMORL-TCTO}
\end{algorithm}

\begin{algorithm}[htb]
	\small
	\caption{Multi-task multi-objective PPO (MMPPO)}
	\begin{algorithmic}[1]
		\REQUIRE
		task set $\Omega$, number of iterations $\Phi$.
		\STATE Initialize offspring population $\mathcal{P}^\prime = \emptyset$;
		\FOR{$\Gamma_i = \langle \mathbf{w}_i, \pi_{\theta_i}, \pi_{\theta_i^{\rm o}}, \mathbf{V}_{\pi_{\theta_i}} \rangle \in \Omega$} 
		
		\FOR{$j = 1,...,\Phi$} 
		\STATE Collect a set of trajectories using sample policy $\pi_{\theta_i^{\rm o}}$;
		\STATE Calculate the advantage function $\mathbf{A}_t$ by Eq. (\ref{Eq: GAE});
		\STATE Calculate the extended advantage function $A^{\mathbf{w}_i}_t = \mathbf{w}_i \mathbf{A}_t$;
		\STATE Update the target policy network's  parameter $\theta_i$ by \\Eq.  (\ref{Eq: surr}) for several epochs;
		\STATE Update the sample policy network’s parameter $\theta_i^{\rm o}$, \\ i.e., $\theta_i^{\rm o}\leftarrow \theta_i $;
		\STATE Update the value network $\mathbf{V}_{\pi_{\theta_i}}$ by Eq. (\ref{Eq: vaule fun});
		\STATE Store the new task $\Gamma_j = \langle \mathbf{w}_i, \pi_{\theta_i}, \pi_{\theta_i^{\rm o}}, \mathbf{V}_{\pi_{\theta_i}} \rangle$ in $\mathcal{P^\prime}$;
		
		\ENDFOR
		\ENDFOR
		\ENSURE
		Offspring population $\mathcal{P^\prime}$.
	\end{algorithmic}
	\label{Algorithm: MMPPO}
\end{algorithm}

\begin{algorithm}[htb]
	\small
	\caption{Task population update (TPU)} 
	\begin{algorithmic}[1]
		\REQUIRE 
		task population $\mathcal{P}$, offspring population $\mathcal{P^\prime}$, reference point $\mathbf{Z}_{\rm ref}$, $P_{\rm num}$, and $P_{\rm size}$.
		\STATE Generate $P_{\rm num}$ evenly distributed weight vectors $\{\mathbf w_1,..., \mathbf w_{P_{\rm num}}\}$;
		\STATE Set performance buffer $\mathcal{B}_i = \emptyset, i=1,...,P_{\rm num}$;

		\FOR{$\Gamma = \langle \mathbf{w}, \pi_{\theta}, \pi_{\theta^{\rm o}}, \mathbf{V}_{\pi_\theta} \rangle \in \{\mathcal{P} \cup \mathcal{P^\prime}$\}} 
		\STATE Calculate objective vector $\mathbf{F}(\pi_{\theta})$; 
		\STATE Set $\mathbf{F}_{\rm temp} =\mathbf{F}(\pi_{\theta}) - \mathbf{Z}_{\rm ref}$;
		\STATE Set index $\hat j=\arg\max_{j=1,...,P_{\rm num}} \{\mathbf{w}_j \mathbf{F}_{\rm temp}\}$;
		\STATE Store task $\Gamma$ in $\mathcal{B}_{\hat j}$;
		\STATE Calculate distance between $\mathbf{F}(\pi_{\theta})$ and $\mathbf{Z}_{\rm ref}$;
		
		\IF{$|\mathcal{B}_{\hat j}|>P_{\rm size}$}
		\STATE Sort all tasks in $\mathcal{B}_{\hat j}$ in descending order of their distances;
		\STATE Retain the first $P_{\rm size}$ tasks in $\mathcal{B}_{\hat j}$ ; 
		\ENDIF
		
		\ENDFOR
		\STATE Set new task population $\mathcal{P}_{\rm new} = \{\mathcal{B}_1 \cup,...,\cup \mathcal{B}_{P_{\rm num}}\}$;
		
		\ENSURE
		population $\mathcal{P}_{\rm new}$.
	\end{algorithmic}
	\label{Algorithm: TPU}
\end{algorithm}

The pseudo-code of EMORL-TCTO is shown in Algorithm \ref{Algorithm: EMORL-TCTO}. We elaborate the two stages above in detail.
\subsubsection{Warm-up Stage}
In this stage, $n$ learning tasks are randomly generated. These tasks share the same state space, action space, and reward function but their dynamics may differ. The task generation procedure is described as follows. 

Firstly, the systematic method \cite{li2014evolutionary} is adopted to generate $n$ evenly distributed weight vectors,  $\mathcal{W}=\{\mathbf w_1,..., \mathbf w_n\}$. Each weight vector is sampled from a unit simplex. $n=\binom{m+\delta-1}{m-1}$ points with a uniform spacing of $1/\delta$, are sampled on the simplex for any number of objectives, where $\delta >0 $ is the number of divisions considered along each objective axis. As \cite{deb2014evolutionary} suggests, to obtain intermediate weight vectors within the simplex, we have $\delta>m$. For example, for the TCTO problem with three objectives ($m=3$), if four divisions ($\delta=4$) are considered for each objective axis, $n=\binom{3+4-1}{3-1}=15$ evenly distributed weight vectors are generated. We plot these weights vectors in Fig. \ref{Figure: EDWV}. 

Secondly, $n$ target policy networks, $\{\pi_{\theta_1},...,\pi_{\theta_n}\}$, are randomly initialized. The corresponding sample policy networks, $\{ \pi_{\theta_1^{\rm o}} ,...,\pi_{\theta_n^{\rm o}} \}$, are initialized, with their parameters set the same as the target policy networks', i.e., $\theta_i^{\rm o} = \theta_i, i=1,...,n$. Then, $n$ multi-objective value networks, $\{\mathbf{V}_{\pi_{\theta_1}},...,\mathbf{V}_{\pi_{\theta_n}}\}$, are randomly initialized. In each value network, the number of neurons in the output layer is the same as that of optimization objectives, i.e., $m$. 

Finally, we denote the set of learning tasks by $\Omega = \{\Gamma_1,...,\Gamma_n\}$, where $\Gamma_i= \langle \mathbf{w}_i, \pi_{\theta_i}, \pi_{\theta_i^{\rm o}}, \mathbf{V}_{\pi_{\theta_i}} \rangle$. After generating the tasks, we run MMPPO to obtain the offspring population, as shown in Algorithm \ref{Algorithm: MMPPO}, where each learning task $\Gamma_i \in \Omega$ is optimized by executing multi-objective PPO (steps 3-11) for a specified number of iterations, $\Phi$ (equals to $\Phi_{\rm warm}$ in this stage).

It is noted that the original MMPPO \cite{xu2020prediction} only stores the last learning task in the offspring population $\mathcal{P^\prime}$ after $\Phi$ iterations, which may throw away promising learning tasks. To avoid the problem, we improve the original MMPPO by storing the new learning task in $\mathcal{P^{\prime}}$ after each iteration. In other words, we save all the learning tasks generated by MMPPO in the offspring population. Thus, running our MMPPO once obtains $n \cdot \Phi$ new learning tasks.  

The warm-up stage can provide a set of promising learning tasks of which policies reside in high-performance region in the search space. To start with these tasks, the EMORL-TCTO's learning process is of low noise, hence more likely to achieve excellent MOO results.

\subsubsection{Evolutionary Stage}
In this stage, the task population $\mathcal{P}$ is first updated based on the offspring population $\mathcal{P^{\prime}}$ (step 9 in Algorithm \ref{Algorithm: EMORL-TCTO}). The task population update procedure is shown in Algorithm \ref{Algorithm: TPU}. We adopt the performance buffer strategy in \cite{xu2020prediction} to update $\mathcal{P}$. A number of performance buffers are used to store $\mathcal{P}$ for the purpose of diversity and performance preservation. Let $P_{\rm num}$ and $P_{\rm size}$ denote the number of performance buffers and their size, respectively. The performance space is evenly divided into $P_{\rm num}$ performance buffers, each of which stores $P_{\rm size}$ learning tasks at most. According to the target policy's objective value, $\mathbf{F}(\pi_\theta)$, and a reference point $\mathbf{Z}_{\rm ref}$, we store the task associated with $\pi_\theta$ in the corresponding performance buffer. 

For an arbitrary performance buffer, we sort the tasks in descending order according to their distances to $\mathbf{Z}_{\rm ref}$. If the number of tasks exceeds $P_{\rm size}$, we only retain the first $P_{\rm size}$ tasks in that buffer. Finally, the learning tasks in all performance buffers form a new task population.

An EP is employed to store nondominated policies found during evolution. In each generation, EP is updated based on the offspring population $\mathcal{P^\prime}$ (step 10 in Algorithm \ref{Algorithm: EMORL-TCTO}). For the target policy $\pi_\theta$ of each learning task in $\mathcal{P^\prime}$, we remove those policies dominated by $\pi_\theta$, and add $\pi_\theta$ to EP if no policies in EP dominates $\pi_\theta$. 

For each weight vector, we select the best learning task from $\mathcal{P}$ and update the set of learning tasks $\Omega$ with it. First of all, we calculate the objective vector $\mathbf{F}(\pi_{\theta_j})$ of the target policy $\pi_{\theta_j}$ of learning task $\Gamma_j, j=1,...,|\mathcal{P}|$, in $\mathcal{P}$. Then, for weight vector $\mathbf{w}_i \in \mathcal{W}$, the best learning task in $\mathcal{P}$ is selected based on $\mathbf{w}_i$ and $\mathbf{F}(\pi_{\theta_j}), j=1,...,|\mathcal{P}|$ (steps 14-16 in Algorithm \ref{Algorithm: EMORL-TCTO}). Finally, the $n$ selected learning tasks are added to $\Omega$. We obtain $\mathcal{P^\prime}$ by running MMPPO with $\Omega$ and $\Phi_{\rm task}$ as its input, where $\Phi_{\rm task}$ is the predefined number of task iterations.

The evolutionary stage terminates when a predefined number of evolution generations are completed. All nondominated policies stored in EP are output as the approximated Pareto optimal policies for the TCTO problem. These policies correspond to different tradeoffs between delay, energy consumption and number of tasks, being helpful for decision makers to compromise between conflicting issues/concerns when designing complicated UAV-assisted MEC systems.

\section{Simulation Results and Discussion}
\label{simulation results}
We first introduce the parameter settings for the UAV-assisted MEC scenario. Assume that the UAV's mission period is 5 minutes and each time slot lasts for 1 second. Therefore, there are $T=300$ time slots. The side lengths of the rectangular area, $x_{\rm max}$ and $y_{\rm max}$, are both set to 400 m. At the beginning of each mission, the UAV takes off at a random position in the rectangular area. In each time slot, the UAV's maximal flying velocity $v_{\rm max}$ and distance $d_{\rm max}$ are set to 30 m/s and 30 m, respectively. The input data size of a computation task, $\alpha$, and the number of CPU cycles required to execute the task, $\beta$, are set to 5 MB and $10^9$ cycles, respectively. For each SD, its parameter of Bernoulli random variable is randomly selected from set $\{0.3, 0.5, 0.7\}$. As for the parameters of pathloss, we set $A_0$, $B_0$, $\theta_0$, $C_0$, and $\eta_0$ to $3.04$, $-23.29$, $-3.61$, $4.14$, and $20.7$, respectively \cite{zhou2020deep}. 

We then describe the parameter settings associated with RL. The number of learning tasks $n$ is set to 15. Each task is associated with a weight vector. So, there are 15 weight vectors, as shown in Fig. \ref{Figure: EDWV}. For each learning task, there are two fully connected layers in the target policy network. Each layer has 64 neurons, with tanh as activation function. The target policy network's output layer uses the sigmoid function to bound actions. Except for the input and output layers, the multi-objective value network shares the same structure and activation function with the target policy network. We use Adam optimizer with a learning rate of 0.0001 to update neural networks. Other parameter configurations are summarized in Table \ref{Table: simulation parameter}.

We finally introduce the test instances. We consider the number of SDs, $K$, and the UAV's flying altitude, $H$, as two important parameters. We specify $K\in \{60, 100, 140\}$ and $H\in \{30,50\}$ and generate six test instances with different combinations of $K$ and $H$. These test instances are listed in Table \ref{Table: test instance}.

\begin{table}
	\centering
	\caption{Parameter configurations in experiments}
	\begin{tabular}{m{6.2cm}<{\raggedleft} m{1.8cm}<{\raggedright}}  
		\hline 
		Parameter & Value \\  
		\hline
		\multicolumn{2}{c}{ Value used in system model} \\
		\hline
		Rotor disc area $(A)$ & 0.503 m$^2$\\
		Fuselage drag ratio $(d_0)$ & 0.6\\
		Maximal distance the UAV can move $(d_{\rm max})$ & 30 m\\
		
		Computing capability of the UAV $(f_{\rm U})$ & 1 GHz\\
		Rotor solidity $(g)$ & 0.05\\ 
		
		Maximum number of tasks in the computing queue $ (N_{\rm max})$ &10\\ 
		Blade profile power $(P_1)$ & 79.86\\
		Induced power $(P_2)$ & 88.63\\
		
		Transmission power of the UAV $(P_{\rm U})$	& 1 W\\
		Tip speed of rotor blade $(U_{\rm tip})$ & 120 m/s \\
		Mean rotor induced velocity in hover $(v_0)$ & 4.03\\
		
		Maximum flying velocity of the UAV $(v_{\rm max})$ & 30 m/s\\
		Channel bandwidth $(W)$ & 10 MHz\\
		Air density $(\rho)$ & 1.225 km/m$^3$\\
		Maximal azimuth angle $(\theta_{\rm max})$ & $\pi/4$\\
		Effective capacitance coefficient $(\kappa)$	& $10^{-26}$\\
		Background noise power $(\sigma^2)$ & $10^{-6}$ W\\
		
		\hline
		\multicolumn{2}{c}{Value used in reinforcement learning}\\
		\hline
		Number of maximum evolution generations $(G_{\rm max})$      & 100\\
		Number of the performance buffers $(P_{\rm num})$ & 200 \\
		Size of each performance buffer $(P_{\rm size})$  & 2\\
		Discount factor $(\gamma)$ & 0.995\\
		Clipping parameter $(\epsilon)$ & 0.2 \\
		Parameter of general advantage estimator $(\lambda)$ & 0.95 \\
		Number of warm-up iterations $(\Phi_{\rm warm})$  & 60\\
		Number of task iterations $(\Phi_{\rm task})$     & 10\\
		Number of divisions of weight vectors $(\delta)$   & 4\\ 		
		\hline  
	\end{tabular}
	\label{Table: simulation parameter}
\end{table}

\begin{table}
	\centering
	\caption{Test instance}
	
	\begin{tabular}{m{2cm}<{\centering} m{2.8cm}<{\centering}m{2.8cm}<{\centering}} 
		\cline{1-3}
		Instance ($K, H$) & Number of SDs ($K$)& Flying altitude ($H$) \\  
		\cline{1-3}
		I-(60,30) & 60 & 30   \\
		I-(60,50)  & 60 & 50 \\
		I-(100,30) & 100 & 30 \\
		I-(100,50) & 100 & 50   \\
		I-(140,30) & 140 & 30  \\
		I-(140,50) & 140 & 50 \\
		
		\cline{1-3}
	\end{tabular}
	\label{Table: test instance}
\end{table}

\subsection{Performance Measure}
We adopt four widely used evaluation metrics to evaluate the performance of EMORL-TCTO, including the inverted generational distance \cite{xu2020bi} , hyper volume \cite{xu2020prediction}, and comprehensive objective indicator \cite{song2022offloading}, and Friedman test \cite{song2020multiobjective}.

\subsubsection{Inverted Generational Distance (IGD)} Let $\mathcal{F}_{\rm true}$ and $\mathcal{F}_{\rm app}$ denote the ture Pareto front and approximated Pareto front found by an MOO algorithm, respectively. IGD is the average distance from each point $v$ in $\mathcal{F}_{\rm true}$ to its nearest counterpart in $\mathcal{F}_{\rm app}$, which is defined as
\begin{equation}
	IGD = \frac{\sum_{v \in \mathcal{F}_{\rm true}}  d(v, \mathcal{F}_{\rm app})}{|\mathcal{F}_{\rm true}|},
\end{equation}
where $d(v, \mathcal{F}_{\rm app})$ is the Euclidean distance between $v$ in $\mathcal{F}_{\rm true}$ and its nearest point in $\mathcal{F}_{\rm app}$. IGD can reflect both the convergence and diversity of an approximated Pareto front. An algorithm with a smaller IGD has better performance. 

Note that we may not know $\mathcal{F}_{\rm true}$ when addressing highly complicated MOO problems, like the TCTO problem. In this case, we collect the best-so-far policies found by all algorithms and select those nondominated from them to mimic the ture Pareto optimal set. We regard the corresponding Pareto front as $\mathcal{F}_{\rm true}$. This method has been widely used when evaluating MOO algorithms in the literature \cite{song2020multiobjective, xu2020bi}.  

\subsubsection{Hyper Volume (HV)}  Let $\mathbf{Z}_{\rm ref} \in \mathbb{R}^m$ be the reference point. HV is defined as
\begin{equation}
	HV = \int_{\mathbb{R}^m} \mathds{1}_{H(\mathcal{F}_{\rm app}) (z)dz},
\end{equation}
where $H(\mathcal{F}_{\rm app})=\{\mathbf{z} | \exists 1\leq j \leq |\mathcal{F}_{\rm app}|:   \mathbf{Z}_{\rm ref} \prec \mathbf{z} \prec \mathbf{Z}_j \}$. $\mathbf{Z}_j$ is the $j$-th point in $\mathcal{F}_{\rm app}$, and $\mathds{1}_{H(\mathcal{F}_{\rm app})}$ is a Dirac delta function that equals 1 if $\textbf{z} \in H(\mathcal{F}_{\rm app})$ and 0, otherwise. 

The HV metric can measure both the convergence and uniformity of an approximated Pareto front without the true Pareto front known in advance. A larger HV value indicates the corresponding algorithm has better performance. In this paper, we set $\mathbf{Z}_{\rm ref}$ to the all-zero vector.

Note that before calculating IGD and HV, we normalize the approximated Pareto front via the Min-Max normalization method. 

\subsubsection{Comprehensive Objective Indicator (COI)} Since the TCTO problem has three objectives, we devise a comprehensive indicator to reflect an MOO algorithm's overall performance, with the task delay, energy consumption, and number of tasks collected taken into account. For each objective vector, we aggregate its objective values into a COI value using the weighted sum method.

Let $\textbf{F}_j=(f_j^1,f_j^2,f_j^3)$ be the $j$-th objective vector in $\mathcal{F}_{\rm app}$. Give a weight vector $\textbf{w}_i=(w_i^1, w_i^2, w_i^3) \in \mathcal{W}$, we define the COI value of $\textbf{F}_j$ as 
\begin{equation}
	COI_j (\textbf{w}_i)= \textbf{w}_i \cdot \textbf{F}_j=\sum_{l=1}^3 w_i^l \cdot f_j^l.
	\label{Eq: COI}
\end{equation}
Based on the COI values, we obtain the best objective vector in $\mathcal{F}_{\rm app}$ associated with $\textbf{w}_i$, which is defined as 
\begin{align}
		&\textbf{F}_b({\textbf{w}_i})=(f_b^1({\textbf{w}_i}),f_b^2({\textbf{w}_i}),f_b^3({\textbf{w}_i})), \nonumber \\
		&b = \mathop{\mathrm{argmax}}\limits_{j=1,...,|\mathcal{F}_{\rm app}|} COI_j (\textbf{w}_i),      
\label{Eq: best COI}
\end{align}
where $f_b^1({\textbf{w}_i}),f_b^2({\textbf{w}_i})$, and $f_b^3({\textbf{w}_i})$ are the best objective values corresponding to $D_{\rm total}, E_{\rm total}$, and $N_{\rm total}$, respectively. According to Eqs. (\ref{Eq: COI}) and (\ref{Eq: best COI}), we obtain the best objective vector for each weight vector in $\mathcal{W}$. After that, we calculate the average task delay (ATD), average energy consumption (AEC), average task number (ATN), and average COI (ACOI), defined as
\begin{align}
	& ATD = \frac{1}{n}  \sum_{i=1}^n f_b^1({\textbf{w}_i}),\\
	& AEC = \frac{1}{n}  \sum_{i=1}^n f_b^2({\textbf{w}_i}),\\
	& ATN = \frac{1}{n}  \sum_{i=1}^n f_b^3({\textbf{w}_i}),\\
	& ACOI = \frac{1}{n} \sum_{i=1}^n COI_b (\textbf{w}_i).
\end{align}

\subsubsection{Friedman Test} The Friedman test, a non-parametric test \cite{friedman1940comparison}, is adopted to measure the differences among MOO algorithms in terms of ATD, AEC, ATN, and ACOI. All algorithms for comparison are ranked, and the average rank scores assigned to them clearly reflect how well they perform.

\subsection{Performance Evaluation}
To thoroughly study the performance of EMORL-TCTO, we implement four baseline algorithms for comparison, including two multi-objective evolutionary algorithms (MOEA), i.e., NSGA-II and MOEA/D, and two multi-policy MORLs, i.e., EDDPG and ETD3. The compared algorithms are described as below. 
\begin{itemize}
	\item NSGA-II: The fast and elitist nondominated sorting genetic algorithm \cite{cui2018joint} adopted to minimize the average task delay and average energy consumption. The population size and number of generations are both set to 100. The crossover and mutation probabilities are set to 0.8 and 0.3, respectively. 
	\item MOEA/D: The multi-objective evolutionary algorithm based on decomposition \cite{song2020multiobjective} used to minimize the average application completion time and average energy consumption. Both the population size and number of generations are set to 100. The number of neighbors for each subproblem is set to 10. 
	\item EDDPG: The evolutionary DDPG, a variant of EMORL-TCTO that uses a multi-task multi-objective DDPG (MMDDPG) instead of MMPPO, i.e., Algorithm \ref{Algorithm: MMPPO}. Note that MMDDPG is extended from the single-policy DDPG \cite{yu2021multi}. We develop EDDPG for performance evaluation purpose. 
	\item ETD3: The evolutionary TD3, another variant of EMORL-TCTO that adopts a multi-task multi-objective TD3 (MMTD3) instead of MMPPO. Note that MMTD3 is extended from the single-policy TD3 \cite{fujimoto2018addressing}. We develop ETD3 for performance evaluation purpose.
	\item EMORL-TCTO: The proposed algorithm in this paper.
\end{itemize}

In NSGA-II and MOEA/D, each gene in a chromosome represents a trajectory control and task offloading decision in a time slot. For fair comparison, EMORL-TCTO, EDDPG, and ETD3 use the same parameter settings.

\begin{figure}
	\centering
	\includegraphics[scale=0.7]{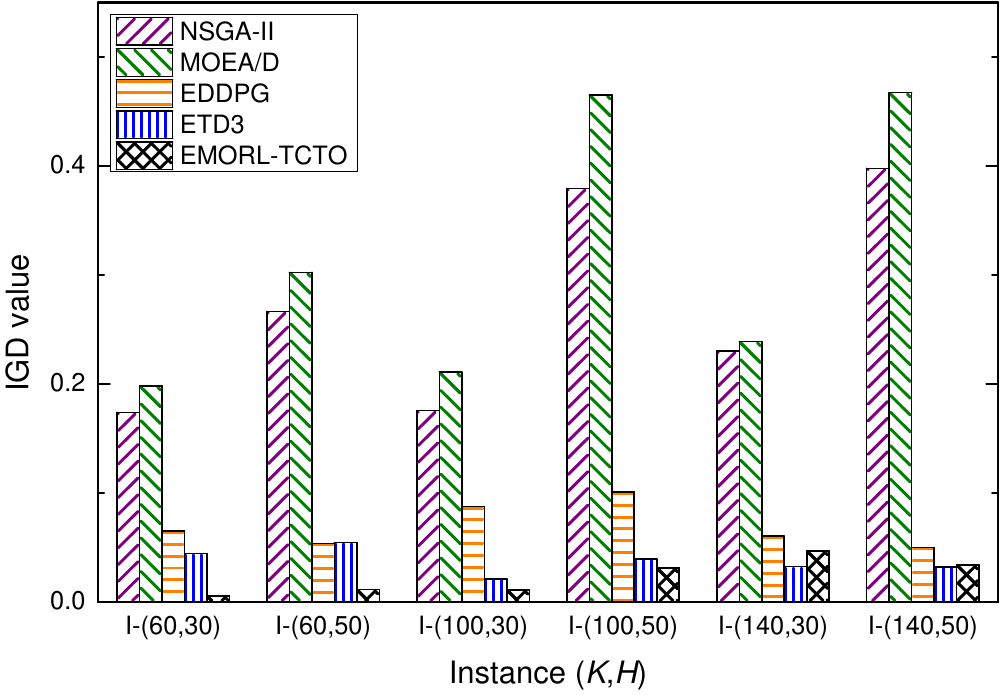} 
	\caption{Results of IGD.}
	\label{Figure: IGD}
\end{figure}

\begin{figure}
	\centering
	\includegraphics[scale=0.7]{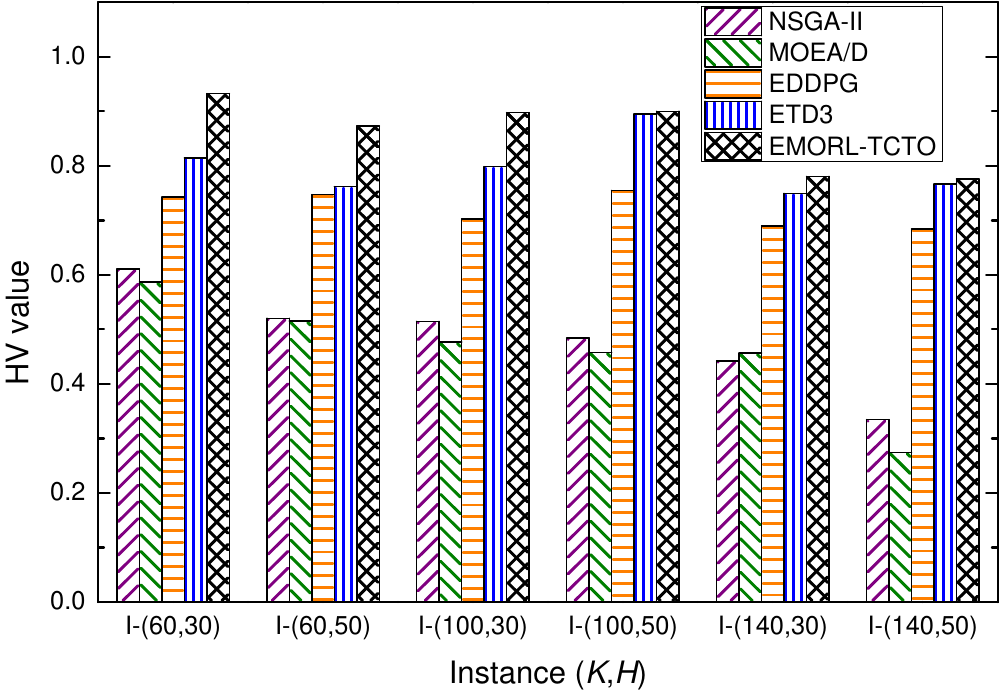} 
	\caption{Results of HV.}
	\label{Figure: HV}
\end{figure}

The results of IGD and HV are shown in Figs. \ref{Figure: IGD} and \ref{Figure: HV}, respectively. First, one can infer that TCTO is a complicated MOO problem since NSGA-II and MOEA/D, both widely recognized, are the two worst algorithms and cannot find a decent Pareto front in all test instances. It is time-consuming that an MOEA with a large encoding length (i.e., 900) results in acceptable nondominated policies. Unfortunately, as the UAV-assisted MEC environment is highly dynamic and full of uncertainty, MOEAs usually do not have sufficient time to converge. That is why NSGA-II and MOEA/D achieve unsatisfied performance. On the other hand, all MORLs outperform NSGA-II and MOEA/D in all test instances. Unlike MOEAs that make decisions for all time slots using a single chromosome, MORLs make real-time decision in each time slot according to the current environment state. Thus, the problem complexity can be reduced. This is why an MORL is more appropriate to address the TCTO problem than an MOEA. EMORL-TCTO obtains the smallest IGD values in four instances and the largest HV values in six instances, demonstrating its superiority over the other four algorithms. 

\begin{figure}
	\centering
	\subfigure[I-(60,30)]{\includegraphics[scale=0.27]{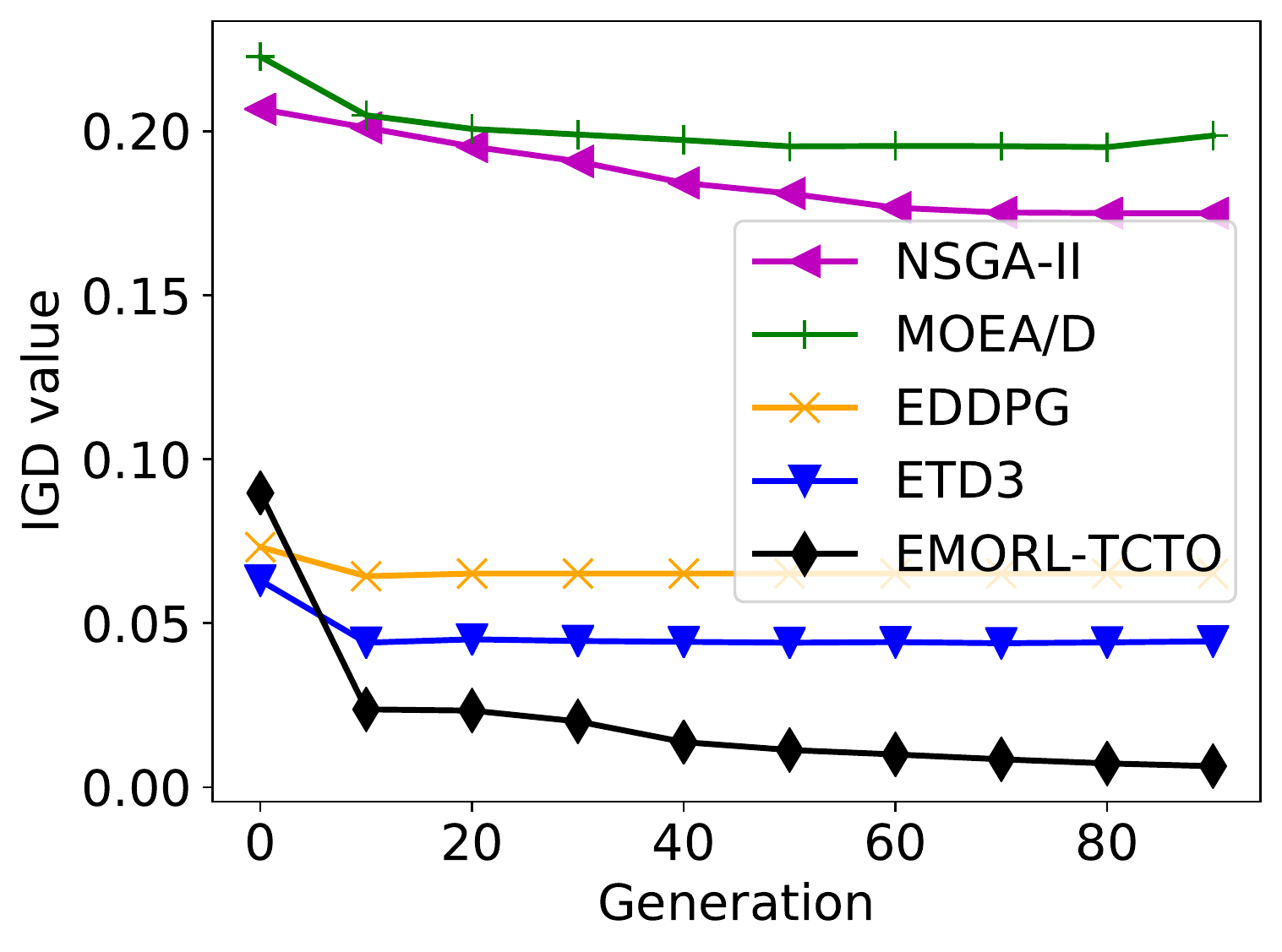}} 
	\subfigure[I-(60,50)]{\includegraphics[scale=0.27]{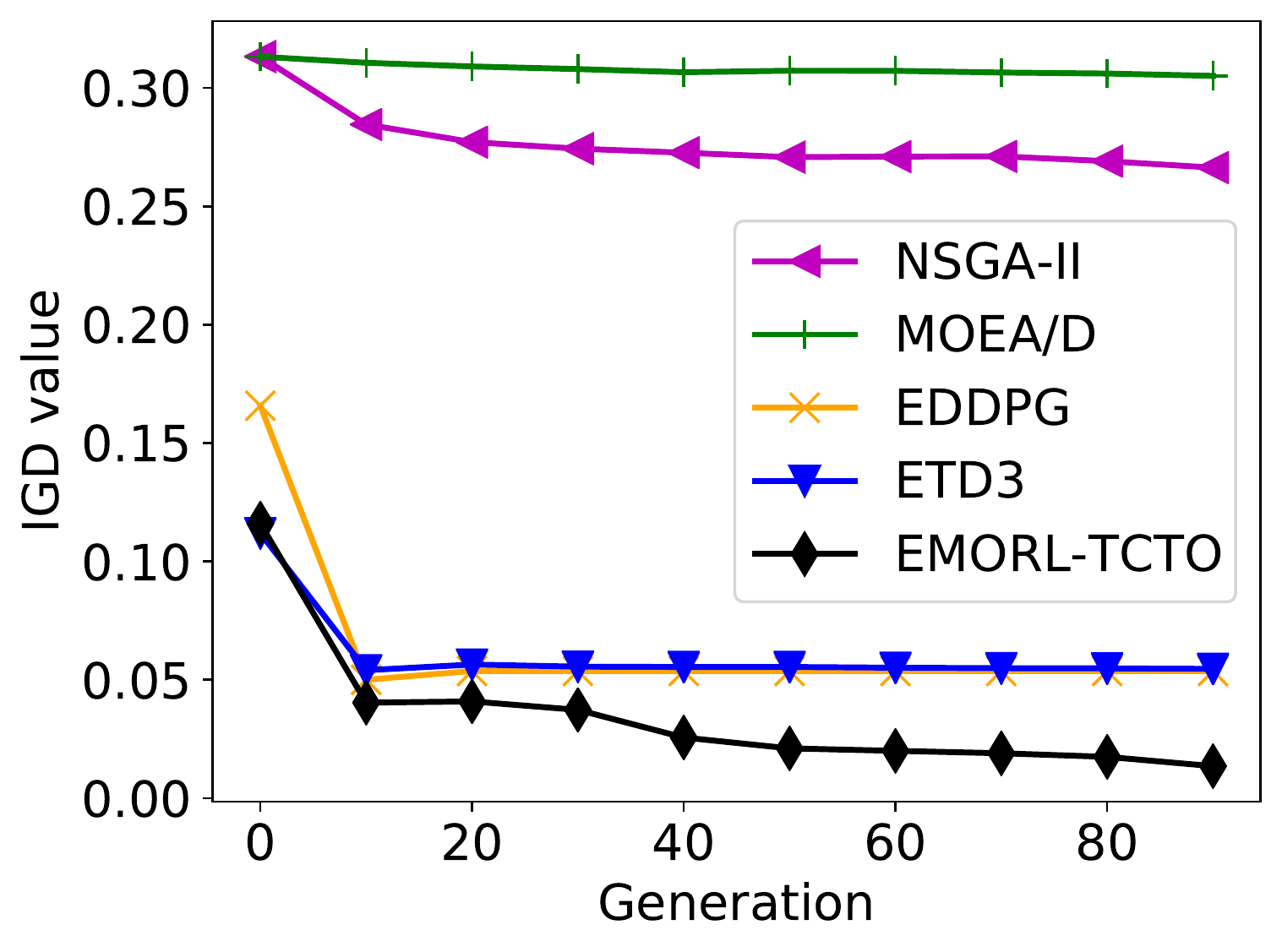}} 
	\subfigure[I-(100,30)]{\includegraphics[scale=0.27]{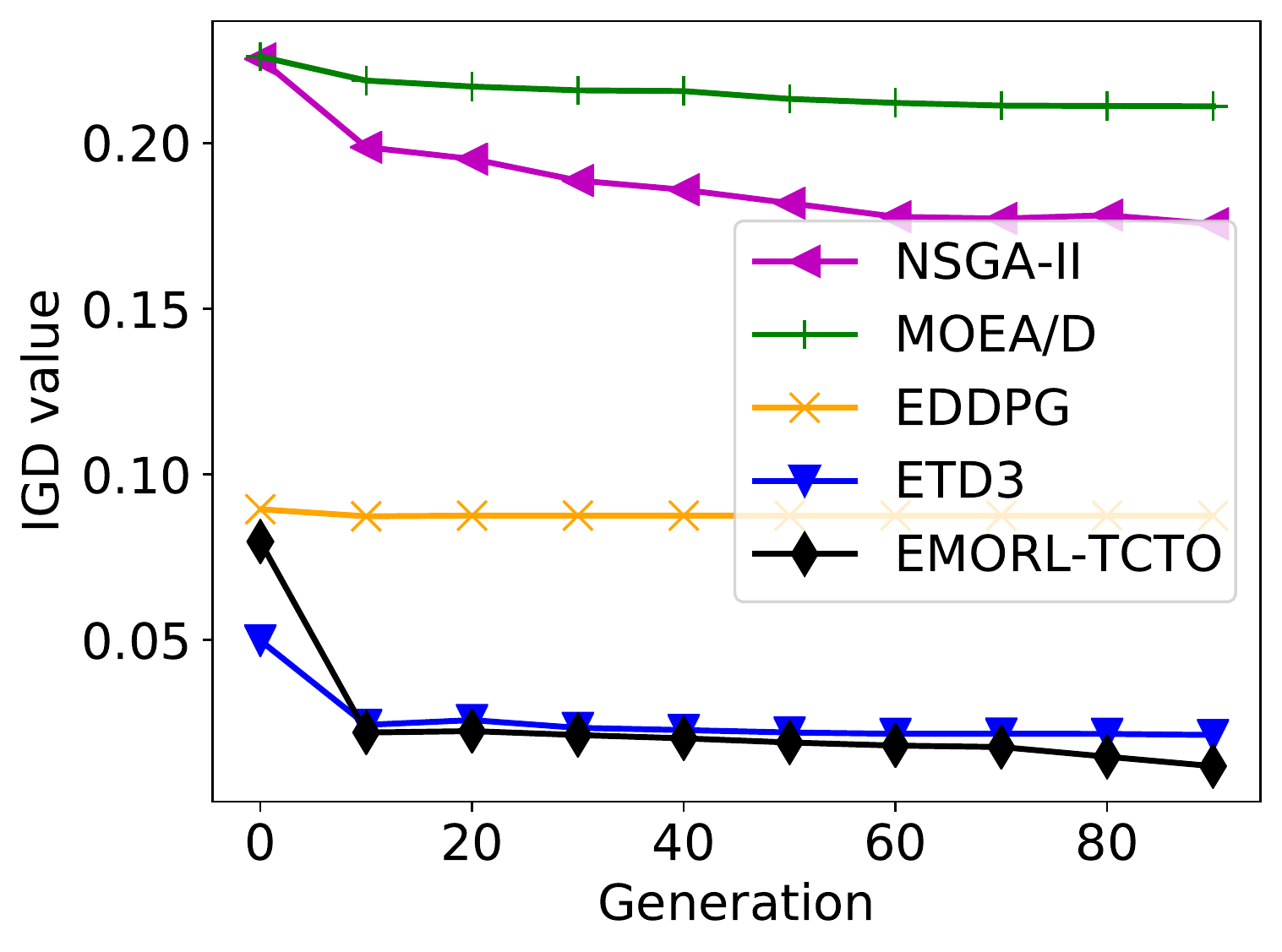}} 
	\subfigure[I-(100,50)]{\includegraphics[scale=0.27]{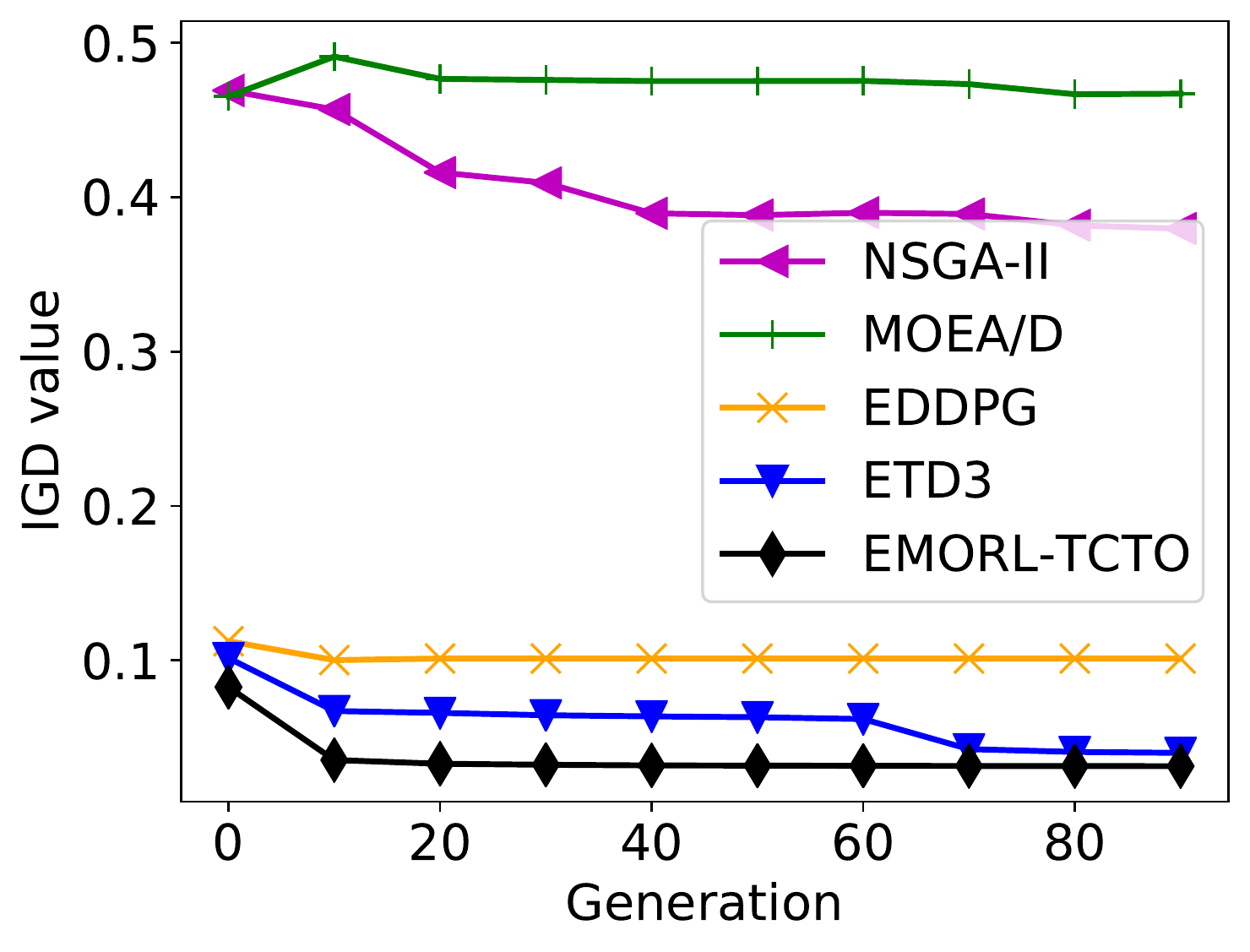}} 
	\subfigure[I-(140,30)]{\includegraphics[scale=0.27]{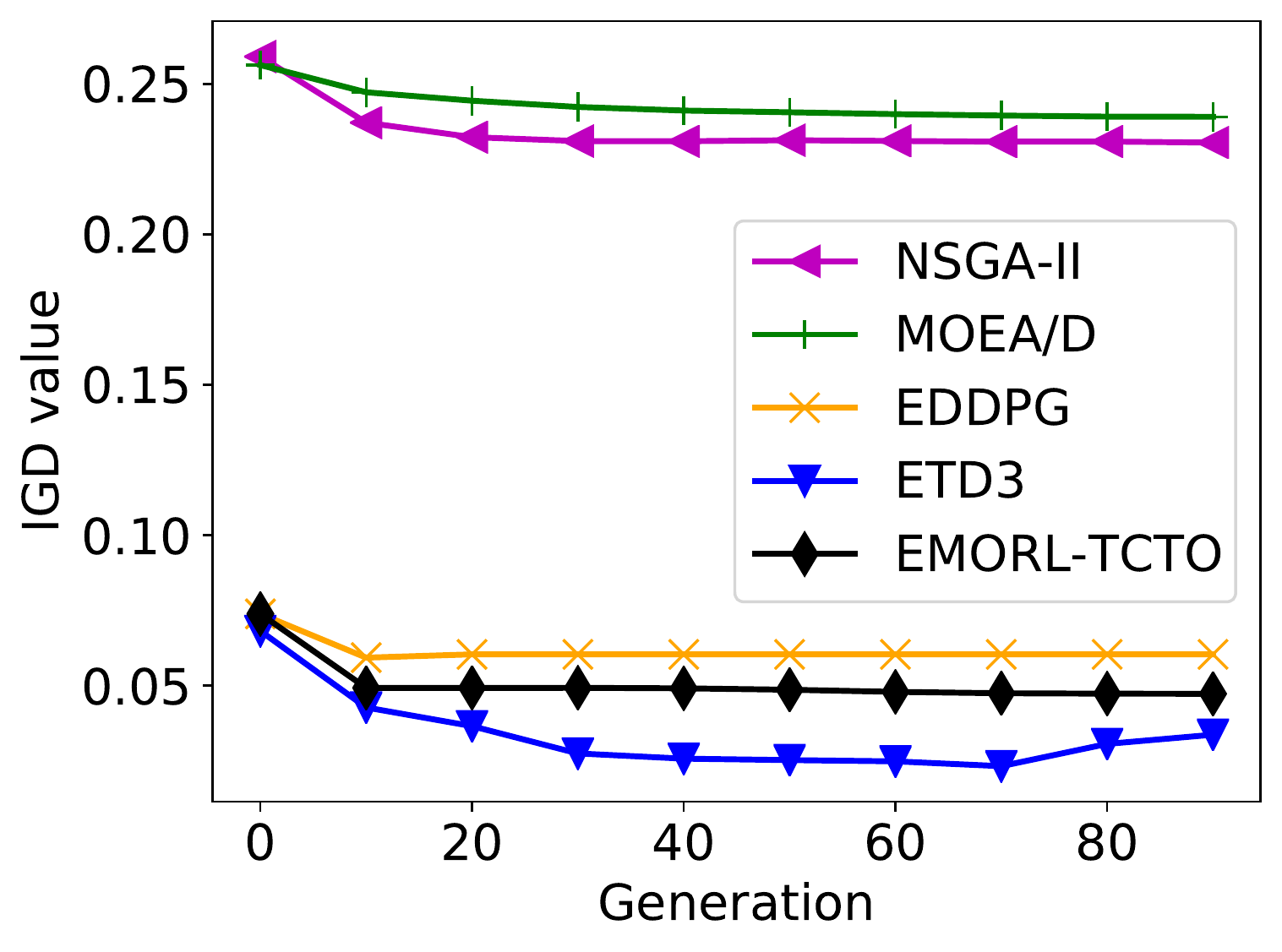}} 
	\subfigure[I-(140,50)]{\includegraphics[scale=0.27]{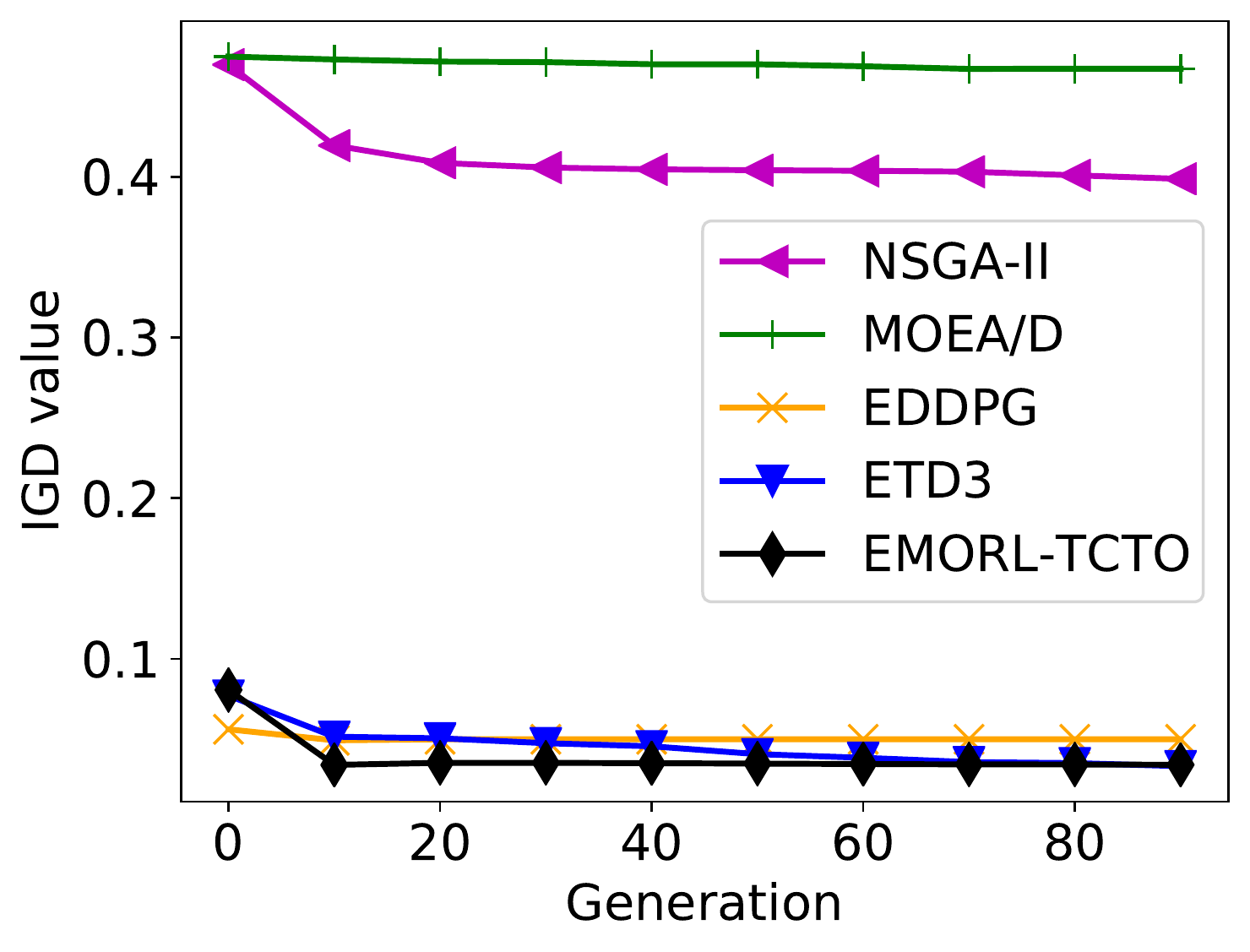}} 
	
	\caption{Convergence curves of five algorithms in terms of IGD.}
	\label{Figure: IGDcurve}
\end{figure}

\begin{figure}
	\centering
	\subfigure[I-(60,30)]{\includegraphics[scale=0.27]{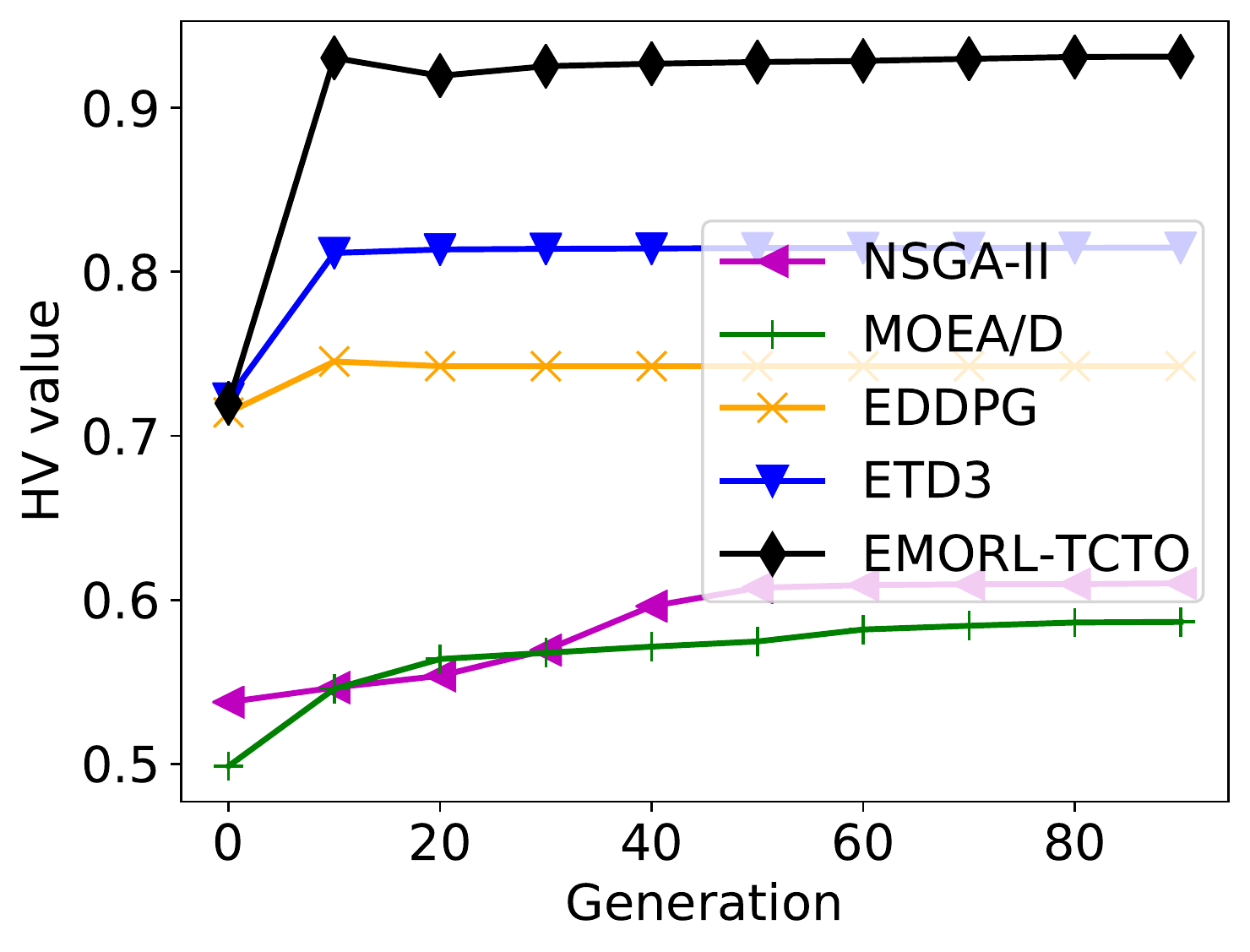}} 
	\subfigure[I-(60,50)]{\includegraphics[scale=0.27]{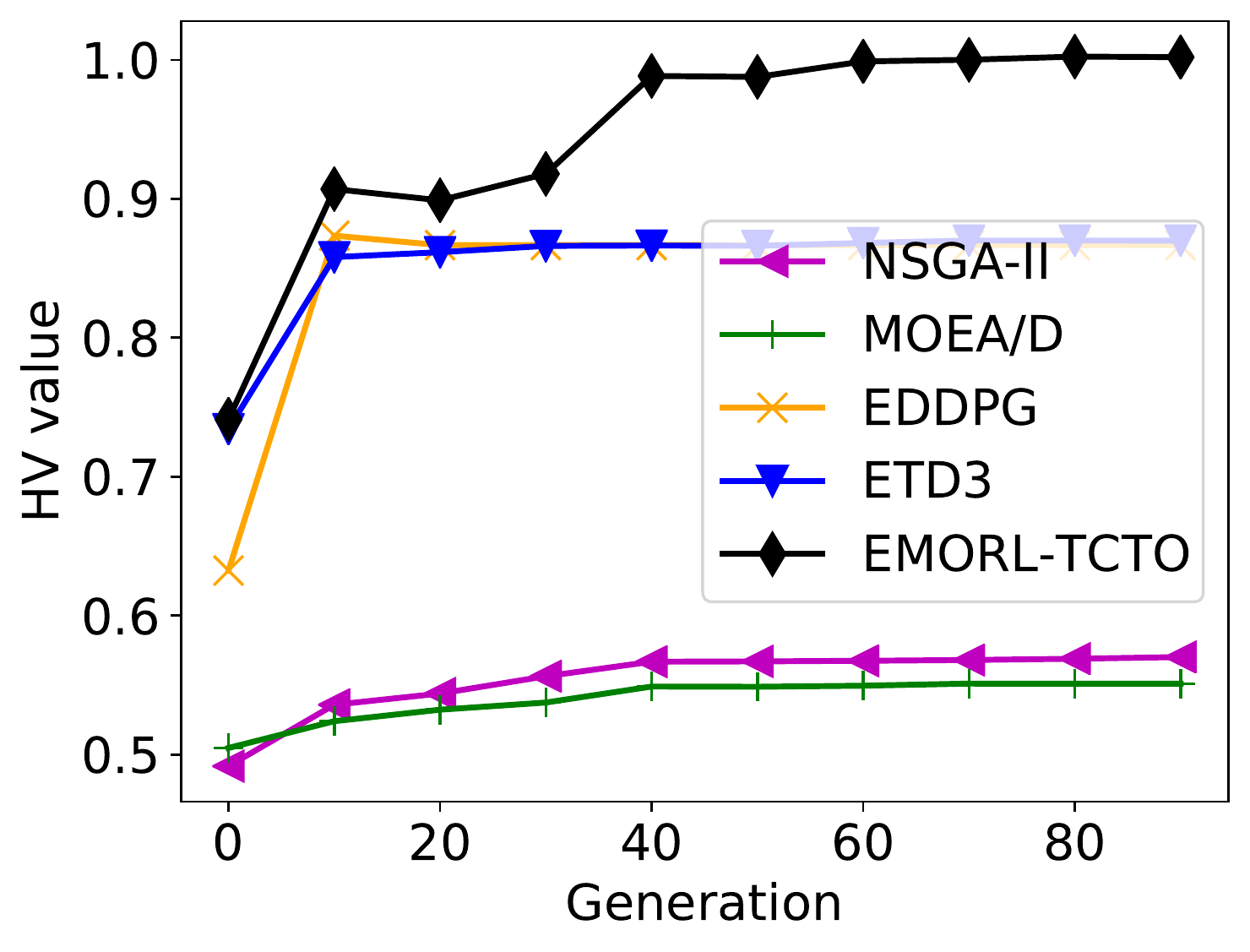}} 
	\subfigure[I-(100,30)]{\includegraphics[scale=0.27]{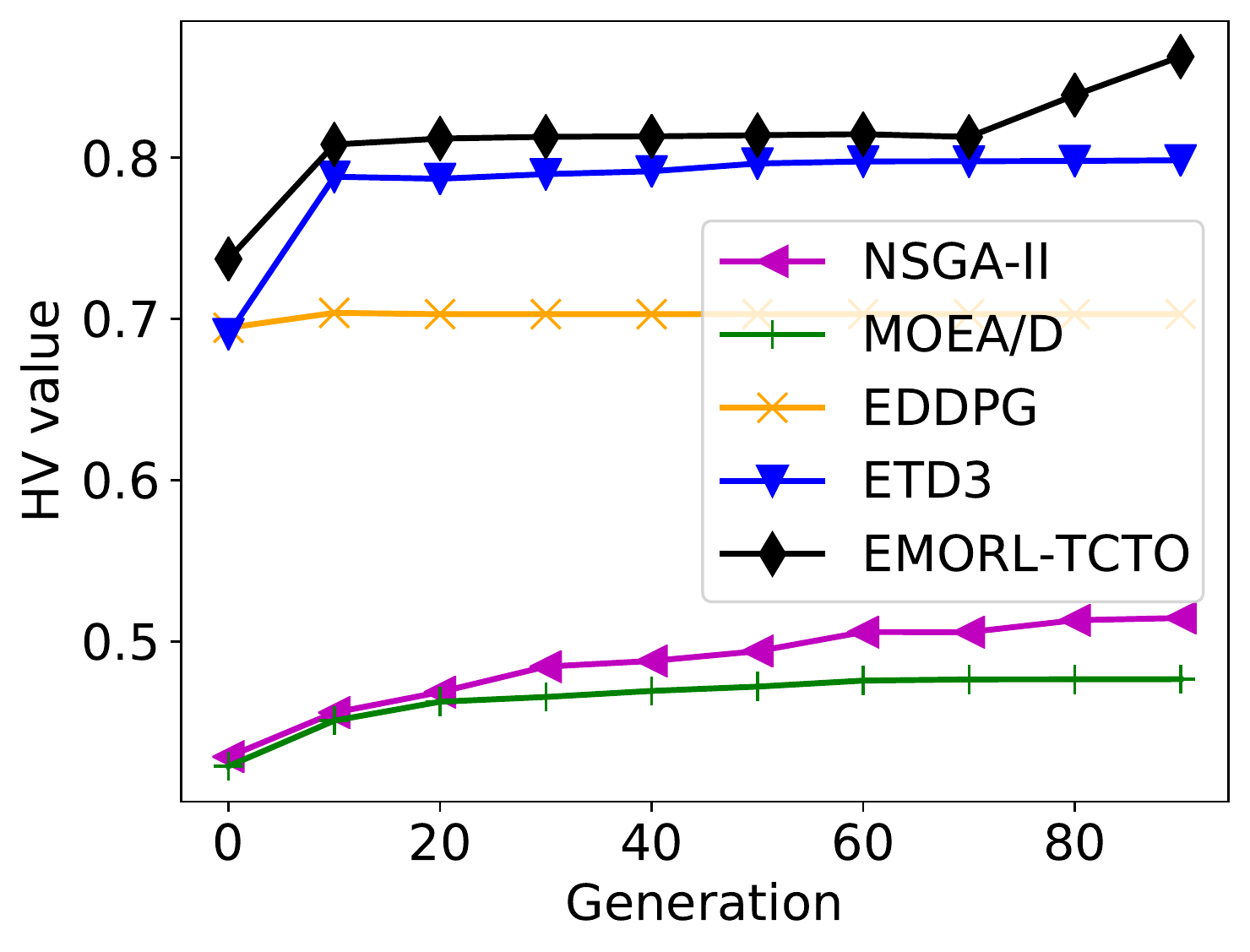}} 
	\subfigure[I-(100,50)]{\includegraphics[scale=0.27]{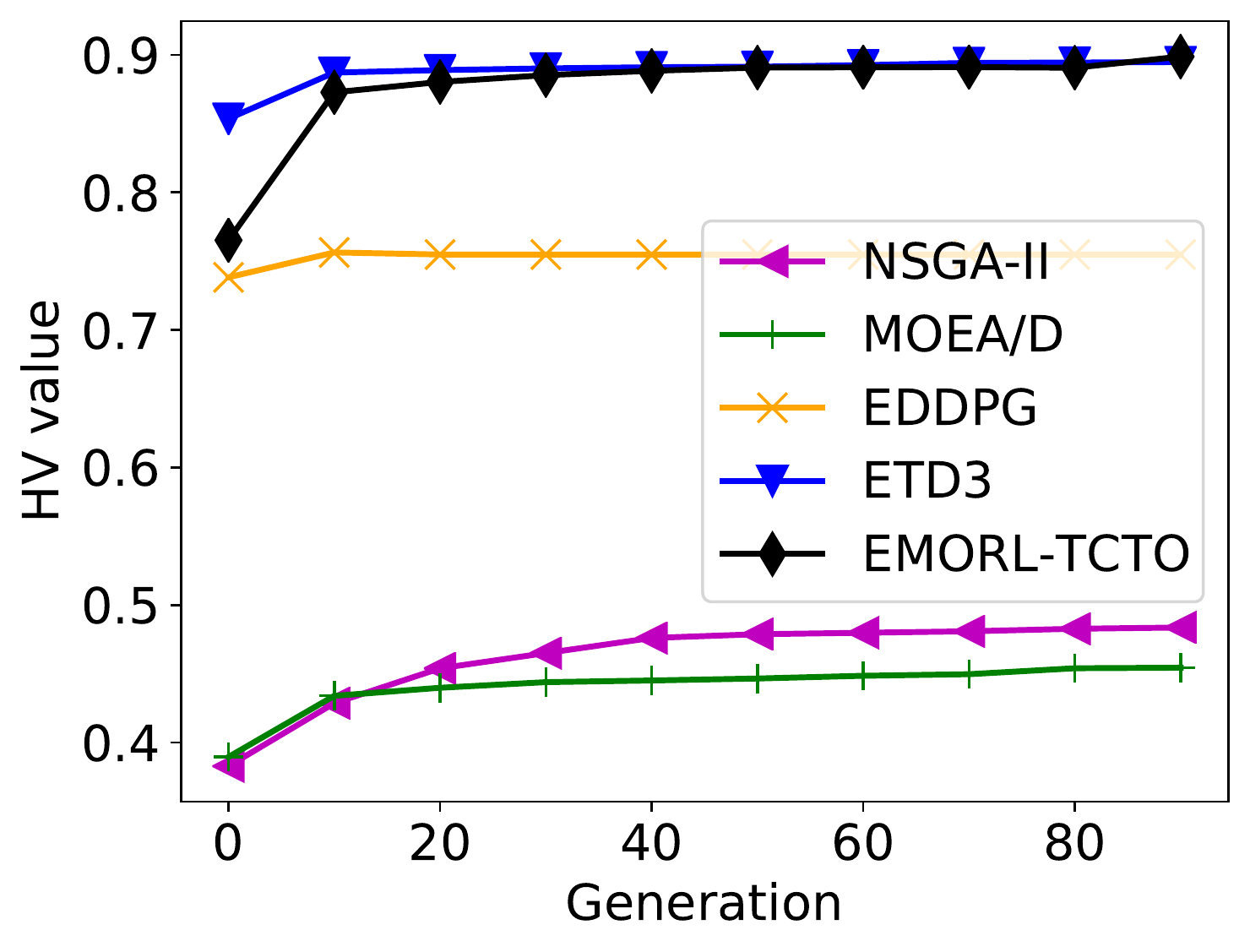}} 
	\subfigure[I-(140,30)]{\includegraphics[scale=0.27]{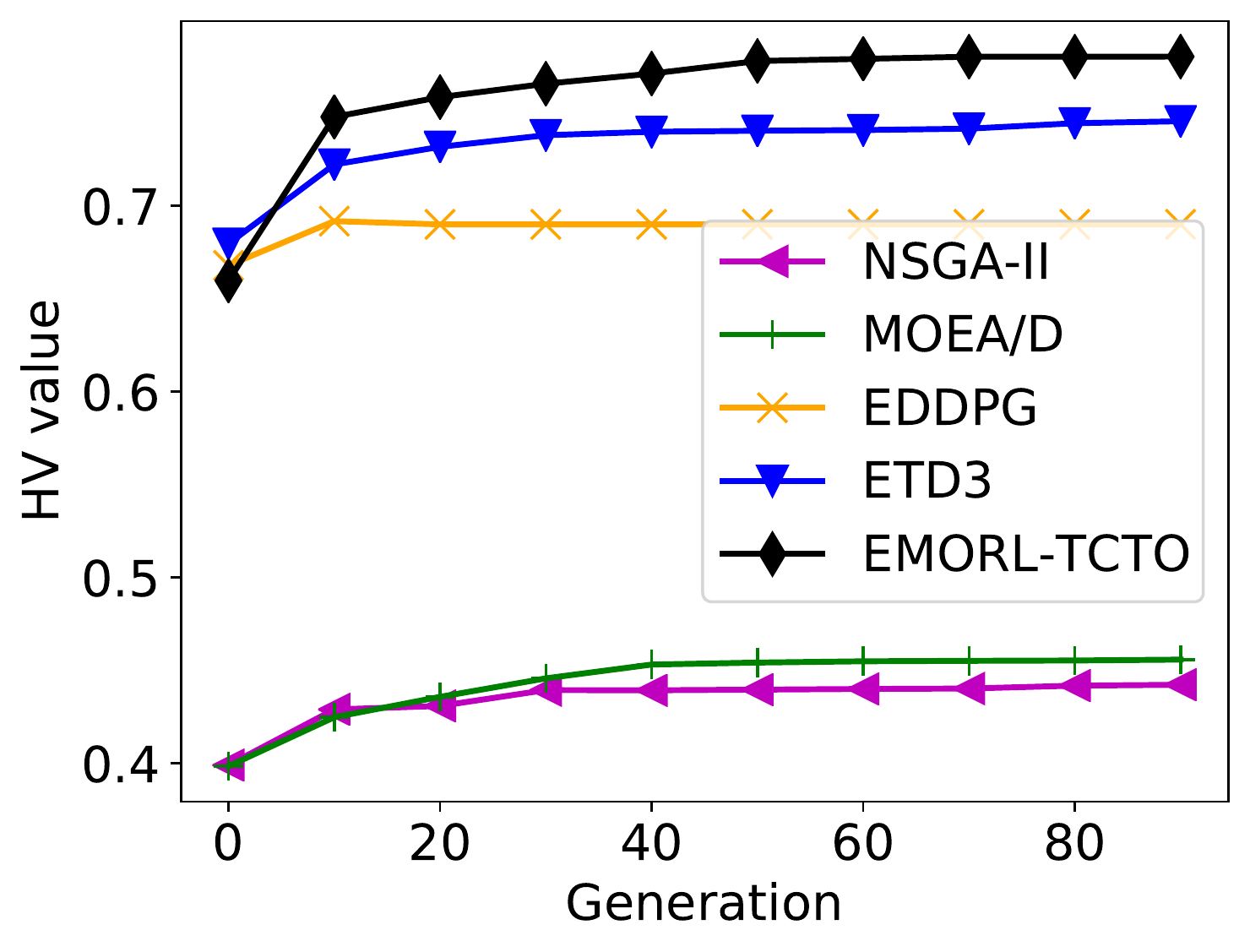}} 
	\subfigure[I-(140,50)]{\includegraphics[scale=0.27]{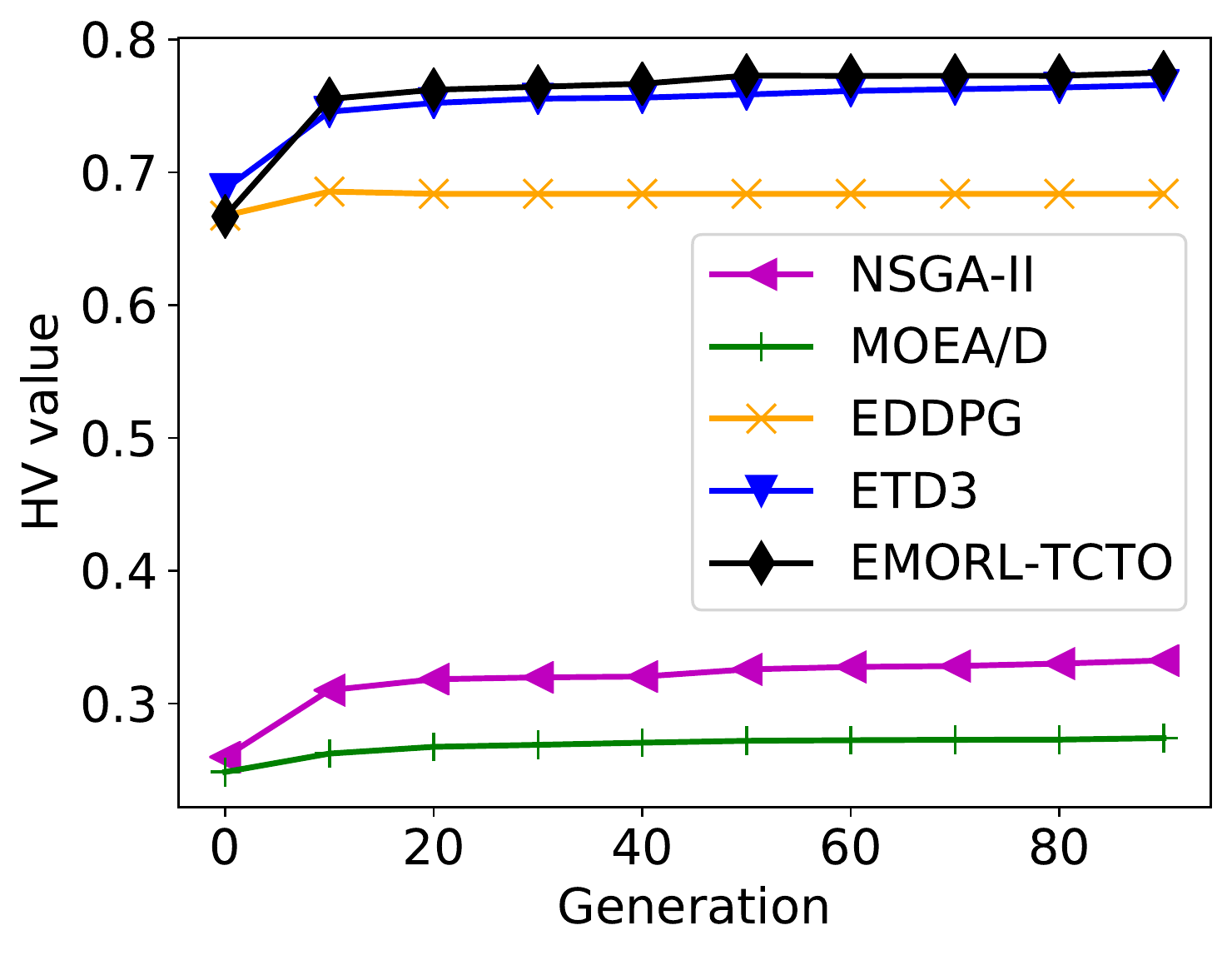}} 
	
	\caption{Convergence curves of five algorithms in terms of HV.}
	\label{Figure: HVcurve}
\end{figure}

To further support our observation above, we plot the convergence curves of IGD and HV obtained by all algorithms in Figs. \ref{Figure: IGDcurve} and \ref{Figure: HVcurve}. It is obvious that EMORL-TCTO is the best among all algorithms in almost all instances except I-(140,30) and I-(140,50) in terms of IGD. Again, NSGA-II and MOEA/D are the two worst algorithms because they easily suffer from rapid diversity loss and premature convergence.

\begin{table*}
	\centering
	\caption{Results of ATD (sec.)}
	
	\begin{tabular}{m{2.33cm}<{\centering} m{2.31cm}<{\centering}m{2.31cm}<{\centering}m{2.31cm}<{\centering}m{2.31cm}<{\centering} m{2.31cm}<{\centering}}     
		\hline
		Instance ($K,H$)&NSGA-II&MOEA/D&EDDPG&ETD3&EMORL-TCTO\\
		\hline
		I-(60,30)&368.1958&445.1122&161.8384&194.0774&\textbf{154.4438}\\
		I-(60,50)&705.8818&913.8073&178.1569&671.9904&\textbf{169.8130}\\
		I-(100,30)&681.5308&782.6583&245.8020&252.0116&\textbf{187.2126}\\
		I-(100,50)&1155.3936&1261.0989&249.7653&\textbf{226.8773}&320.4334\\
		I-(140,30)&933.4857&1104.9209&236.5322&\textbf{236.5244}&592.3761\\
		I-(140,50)&1618.8890&1839.8178&451.0401&640.4673&\textbf{401.9143}\\
		
		\hline
	\end{tabular}
	\label{Table: delay}
\end{table*}

\begin{table*}
	\centering
	\caption{Results of AEC ($\times 100$ J)}
	
	\begin{tabular}{m{2.33cm}<{\centering} m{2.31cm}<{\centering}m{2.31cm}<{\centering}m{2.31cm}<{\centering}m{2.31cm}<{\centering} m{2.31cm}<{\centering}}     
		\hline
		Instance ($K,H$)&NSGA-II&MOEA/D&EDDPG&ETD3&EMORL-TCTO\\
		\hline
		I-(60,30)&541.7004&584.1528&743.8614&622.0636&\textbf{524.3706}\\
		I-(60,50) &573.9115&616.5412&724.3359&762.2767&\textbf{487.8757}\\
		I-(100,30) &574.1068&623.6895&762.4399&\textbf{567.4193}&580.9955\\
		I-(100,50) &679.2424&579.8798&906.7152&737.1444&\textbf{549.9695}\\
		I-(140,30) &\textbf{626.0573}&636.4504&851.3875&776.7553&710.1657\\
		I-(140,50) &677.1887&\textbf{674.9763}&852.3247&917.8795&818.8782\\
		\hline
	\end{tabular}
	\label{Table: energy}
\end{table*}

\begin{table*}
	\centering
	\caption{Results of ATN}
	
	\begin{tabular}{m{2.33cm}<{\centering} m{2.31cm}<{\centering}m{2.31cm}<{\centering}m{2.31cm}<{\centering}m{2.31cm}<{\centering} m{2.31cm}<{\centering}}     
		\hline
		Instance ($K,H$)&NSGA-II&MOEA/D&EDDPG&ETD3&EMORL-TCTO\\
		\hline
		
		I-(60,30)&443.3333&494.6000&623.4000&647.9333&\textbf{679.6000}\\
		I-(60,50)&826.2667&882.7333&1224.2000&1158.6667&\textbf{1306.6667}\\
		I-(100,30)&785.1333&812.2667&1107.3333&1152.7333&\textbf{1320.3333}\\
		I-(100,50)&1476.8000&1650.3333&1914.0000&1919.6667&\textbf{2059.7333}\\
		I-(140,30)&1071.7333&1227.2667&1717.6667&1760.2667&\textbf{1967.7333}\\
		I-(140,50)&2230.6000&2313.2667&3205.7333&3589.2667&\textbf{3767.9333}\\
		
		\hline
	\end{tabular}
	\label{Table: number}
\end{table*}

\begin{table*}
	\centering
	\caption{Results of ACOI}
	
	\begin{tabular}{m{2.33cm}<{\centering} m{2.31cm}<{\centering}m{2.31cm}<{\centering}m{2.31cm}<{\centering}m{2.31cm}<{\centering} m{2.31cm}<{\centering}}    
		\hline
		Instance ($K,H$) &NSGA-II&MOEA/D&EDDPG&ETD3&EMORL-TCTO\\
		\hline
		I-(60,30)&0.3762&32.4460&88.4897&123.2103&\textbf{170.9732}\\
		I-(60,50)&88.1119&51.8222&358.2882&359.2345&\textbf{466.6511}\\
		I-(100,30)&74.9282&5.7390&316.0606&357.1098&\textbf{435.7963}\\
		I-(100,50)&253.9011&210.2697&633.6083&704.2282&\textbf{747.0715}\\
		I-(140,30)&186.4932&169.4821&585.0505&632.3760&\textbf{726.9881}\\
		I-(140,50)&569.7772&482.152&1171.9050&1344.7181&\textbf{1447.5302}\\
		\hline
	\end{tabular}
	\label{Table: COI}
\end{table*}

\begin{table*}[!htp]
	\small
	\centering
	\caption{Rankings of five algorithms}
	\begin{tabular}{m{2.15cm}<{\centering} 
			m{1.2cm}<{\centering} m{1.1cm}<{\centering} m{0.001cm}<{\centering} 
			m{1.2cm}<{\centering} m{1.1cm}<{\centering} m{0.001cm}<{\centering} 
			m{1.2cm}<{\centering} m{1.1cm}<{\centering} m{0.001cm}<{\centering} 
			m{1.2cm}<{\centering} m{1.1cm}<{\centering}}    
		\hline   
		\multirow{2}*{Algorithm}   
		& \multicolumn{3}{c}{ATD} & \multicolumn{3}{c}{AEC} & \multicolumn{3}{c}{ATN} & \multicolumn{2}{c}{ACOI} \\
		
		\cline{2-3} \cline{5-6} \cline{8-9} \cline{11-12}  
		& Average rank & Position && 
		Average rank & Position && 
		Average rank & Position && 
		Average rank & Position \\
		
		\hline
		NSGA-II&	4.0000&	4&&	2.0000&	1&&	5.0000&	5&&	4.0000&	4\\
		MOEA/D&  	5.0000&	5&&	2.5000&	2&&	4.0000&	4&&	5.0000&	5\\
		EDDPG&	    2.0000&	2&&	4.6667&	4&&	2.8333&	3&&	3.0000&	3\\
		ETD3&	    2.3333&	3&&	3.8333&	3&&	2.1667&	2&&	2.0000&	2\\
		EMORL-TCTO&	    1.6667&	1&&	2.0000&	1&&	1.0000&	1&&	1.0000&	1\\

		\hline 
	\end{tabular}
	\label{Table: Friedman test}
\end{table*}
Tables \ref{Table: delay}-\ref{Table: number} show the ATD, AEC, and ATN values obtained by the five algorithms. Note that the best results are in bold. No matter which one gets fixed, $K$ or $H$, the corresponding ATD, AEC, and ATN values tend to grow up as the other increases. Firstly, the larger the number of SDs located in the rectangular area, the more the computation tasks need to be collected by the UAV. Secondly, given that the UAV cannot fly over its maximum allowable altitude, the higher the flying altitude, the larger the UAV's coverage, thus the more the computation tasks can be collected. However, collecting more tasks by the UAV leads to larger task processing delay and higher energy consumption because it has more tasks to handle. Tables \ref{Table: delay}-\ref{Table: number} well support this.

In Table \ref{Table: delay}, it is easily seen that EMORL-TCTO performs better than the other algorithms in four instances except I-(100,50) and I-(140,30). ETD3 achieves the smallest ATD value in I-(100,50) and I-(140,30). However, it is worse than EMORL-TCTO in terms of AEC and ATN, with all instances considered. For instance, although ETD3 obtain the smallest ATD value in I-(100,50) and I-(140,30), its AEC and ATN values are both beaten by EMORL-TCTO's. 

\raggedbottom 
As for the AEC values shown in Table \ref{Table: energy}, EMORL-TCTO outperforms the others in I-(60,30), I-(60,50), and I-(100,50). Although NSGA-II and MOEA/D achieve decent AEC results in I-(140,30) and I-(140,50), they do not perform well regarding ATD and ATN. For example, while NSGA-II obtains the smallest AEC value in I-(140,30), this algorithm causes larger ATD and smaller ATN values than EMORL-TCTO. Similar phenomenon can be observed on MOEA/D. ETD3 obtains the best AEC value in I-(100,30), but its ATD and ATN values are worse than EMORL-TCTO's. 

As shown in Table \ref{Table: number}, EMORL-TCTO is the best as it results in the largest ATN in every instance. It means EMORL-TCTO allows the UAV to collect sufficient number of computation tasks from SDs by appropriately controlling the UAV's flying trajectory, during its entire mission period. 

As aforementioned, the ACOI indicator reflects an MOO algorithm's overall optimization performance. Table \ref{Table: COI} lists the results of ACOI obtained by the five algorithms for comparison. It is easily seen that EMORL-TCTO overweighs NSGA-II, MOEA/D, EDDPG, and ETD3 in all test instances since EMORL-TCTO can better balance between objectives. In addition, the Friedman test is adopted to rank the five algorithms. Based on the ATD, AEC, ATN, and ACOI values, the average rankings and positions of algorithms are calculated and shown in Table \ref{Table: Friedman test}. One can clearly observe that EMORL-TCTO obtains the best overall performance.

\section{Conclusion}
\label{conclusion}
We model the trajectory control and task offloading (TCTO) problem by multi-objective Markov decision process (MOMDP) and adapt a multi-policy multi-objective reinforcement learning algorithm to address the problem. The proposed EMORL-TCTO can output plenty of nondominated policies for various user preferences in each run, clearly reflecting the conflicts between objectives. Compared with NSGA-II, MOEA/D, EDDPG, and ETD3, our algorithm strikes better balance between the objectives in four out of six instances regarding inverted generational distance and in all the six instances regarding hyper volume. EMORL-TCTO is also the best in most instances with respect to system-related metrics, including the average task delay, average UAV's energy consumption, average number of tasks collected by the UAV, and average comprehensive objective indicator. In addition, EMORL-TCTO takes the first position in the Friedman test. Hence, the performance comparison demonstrates EMORL-TCTO's suitability to tackle the TCTO problem and its potential to be applied to multi-objective UAV-assisted MEC scenarios.

\balance
\bibliographystyle{IEEEtran} 

\bibliography{cas-refs.bib}

\end{document}